\documentclass[a4paper,12pt]{iopart}
\usepackage{iopams}
\usepackage{setstack}
\usepackage{graphicx}
\usepackage{bm}
\usepackage{epsfig}

\newcommand{\diag}{{\rm diag\,}}
\newcommand{\sign}{{\rm sign\,}}
\newcommand{\Str}{{\rm Str\,}}
\newcommand{\Sdet}{{\rm Sdet\,}}
\newcommand{\Pf}{{\rm Pf\,}}
\newcommand{\UOSp}{{\rm UOSp\,}}

\newcommand{\U}{{\rm U\,}}

\newcommand{\Herm}{{\rm Herm\,}}
\newcommand{\Ber}{{\rm Ber\,}}
\newcommand{\RE}{{\rm Re\,}}

\newcommand{\eins}{\leavevmode\hbox{\small1\kern-3.8pt\normalsize1}}
\eqnobysec

\begin{document}

\newtheorem{definition}{Definition}[section]
\newtheorem{assumption}[definition]{Assumption}
\newtheorem{theorem}[definition]{Theorem}
\newtheorem{lemma}[definition]{Lemma}
\newtheorem{corollary}[definition]{Corollary}

\title[Determinantal structures]{Derivation of determinantal structures for random matrix ensembles in a new way}
\author{Mario Kieburg$^\dagger$ and Thomas Guhr}
\address{Universit\"at Duisburg-Essen, Lotharstra\ss e 1, 47048 Duisburg, Germany}
\eads{$^\dagger$ \mailto{mario.kieburg@uni-due.de}}

\date{\today}

\begin{abstract}
 There are several methods to treat ensembles of random matrices in symmetric spaces, circular matrices, chiral matrices and others. Orthogonal polynomials and the supersymmetry method are particular powerful techniques. Here, we present a new approach to calculate averages over ratios of characteristic polynomials. At first sight paradoxically, one can coin our approach ``supersymmetry without supersymmetry'' because we use structures from supersymmetry without actually mapping onto superspaces. We address two kinds of integrals which cover a wide range of applications for random matrix ensembles. For probability densities factorizing in the eigenvalues we find determinantal structures in a unifying way. As a new application we derive an expression for the $k$--point correlation function of an arbitrary rotation invariant probability density over the Hermitian matrices in the presence of an external field.
\end{abstract}

accepted for publication in J. Phys. A: Math. Theor.

\section{Introduction}\label{sec1}

Random matrix theory has a wide range of applications \cite{Efe97,GMW98,VerWet00,Meh04} resulting in a large number of different matrix ensembles. To name but a few examples, generic features of Hamilton operators are modeled by ensembles over the symmetric spaces \cite{VWZ85,GMW98,Guh06}. In a relativistic setting, chiral (Laguerre) random matrix ensemble have to be used, involving one \cite{VerWet00,FyoStr02} or two \cite{SWG99,Ake01,Ake02,Ake03,Osb04} matrices. Another example is the statistics of the zero points of the Riemann zeta function \cite{HKC00,KeaSna00,HKC01,BHNY08}, derived by Dyson's circular ensembles \cite{Dys62}.

Averages over ratios of characteristic polynomials play an important role in the investigation of random matrix ensembles. The matrix Green function can be generated by one characteristic polynomial in the denominator and one in the numerator \cite{Zir06,Guh06}. For the calculation of the free energy, one may use the replica trick \cite{VerZir85}. The moments of the Riemann $\zeta$-function are also of interest for number theorists \cite{KeaSna00,HKC01}. Mathematicians are interested in averages over ratios of characteristic polynomials because of the connection to Weyl's character formula \cite{HPZ05,CFZ07}. In models for Quantum Chromodynamics (QCD) \cite{VerWet00,Ake01} and in the analysis of the sign problem \cite{BloWet09}, one employs mean values of characteristic polynomials.

To calculate such mean values, commonly two techniques are used, the method of orthogonal polynomials \cite{BreHik00c,Ber04,Meh04} and the supersymmetry method \cite{Efe83,VWZ85,Guh06,Som07,LSZ07,SVZ08,KGG08,KSG09}. In both methods determinantal structures appear for rotation invariant ensembles over the Hermitian matrices \cite{MehNor01,BDS03,FyoStr03,GGK04,BorStr05}. The probability densities of such ensembles have to factorize in the eigenvalues of the matrices. The determinantal structures extremely simplify the calculation since all $k$--point functions are determined by one--point and two--point correlations. Also for other ensembles such as the non-Hermitian ensembles \cite{AkeVer03,Ber04}, the chiral ensembles \cite{SWG99,FyoStr02,Ake01} and the circular ensembles \cite{BasFor94,CFS05}, determinantal structures were found.

In the orthogonal polynomial method as well as in the supersymmetry method every single ensembles has to be calculated in a particular way. Either one has to find the measure to construct the orthogonal polynomials or one has to identify the superspace corresponding to the ordinary integration domain. We consider two types of integrals related to mean values of ratios of characteristic polynomials. Determinantal structures stemming from supersymmetry, such as those found by Basor and Forrester \cite{BasFor94}, yield determinantal structures of these integrals. Here, we establish the link to supersymmetry. To the best of our knowledge this connection has not been observed before. Our method is based on an algebraic manipulation of the characteristic polynomials and the Jacobian or the Berezinian resulting from changing integration variables. We neither use the Mehta--Mahoux theorem nor a mapping onto superspace. Both types of integrals cover a wide range of applications for unitarily rotation invariant random matrix ensembles and for ensembles over all three rotation groups or their Lie algebras. As a particular example, we consider an intermediate ensemble from arbitrary unitarily invariant ensembles over Hermitian matrices to a rotation invariant ensemble over one of the symmetric spaces. This generalizes known results  \cite{PanMeh83,MehPan83,Guh96b,Guh96c,BreHik98,Sto02,GuhSto04,Guh06b,Joh07}. For this example we use the supersymmetry method. Thereby, we demonstrate that our method works for calculations within superspace, too.

In Sec. \ref{sec1.b}, we give an outline of our approach. We present a determinantal structure of Berezinians resulting from a diagonalization of symmetric supermatrices in Sec. \ref{sec2}. This is then applied to two types of integrals discussed in Sec. \ref{sec3} which yield determinantal structures. In Sec. \ref{sec4}, we present a series of ensembles whose mean values of characteristic polynomial ratios are special cases of one type of integral presented in Sec. \ref{sec3}. In particular, we consider Hermitian matrices  which we investigate in Sec. \ref{sec5} as well. In Sec. \ref{sec5} we, also, study examples for the other type of integral. The main focus is on the Hermitian matrix ensemble with arbitrary unitarily rotation invariant probability density in the presence of an external field. In the appendices, we perform some explicit calculations.

\section{Sketch of the idea: ``Supersymmetry without supersymmetry''}\label{sec1.b}

As a guideline for the reader, we present here the main ideas of our approach for one particular ensemble. We choose $\kappa=\diag(\kappa_{11},\ldots,\kappa_{k1},\kappa_{12},\ldots,\kappa_{k2})=\diag(\kappa_1,\kappa_2)$ in such a way that the integrals below are well defined. For many applications such as for Hermitian matrix ensembles one considers averages over ratios of characteristic polynomials
\begin{equation}\label{1.1}
 \displaystyle Z(\kappa)=\int P(H)\prod\limits_{j=1}^{k}\frac{\det(H-\kappa_{j2}\eins_N)}{\det(H-\kappa_{j1}\eins_N)}d[H]\,.
\end{equation}
Here, $\eins_N$ is the $N\times N$ unit matrix. The probability density $P$ is rotation invariant and factorizes in the eigenvalues of the matrix $H$. We diagonalize $H$ in its eigenvalues $E_1,\ldots,E_N$. The Jacobian is the second power of the Vandermonde determinant $\Delta_N(E)$. We expand one of the Vandermonde determinants and have up to a constant $c$
\begin{equation}\label{1.2}
 \displaystyle Z(\kappa)=c\int\prod\limits_{a=1}^N\left[P(E_a)E_a^{a-1}\prod\limits_{b=1}^{k}\frac{E_a-\kappa_{b2}}{E_a-\kappa_{b1}}\right]\Delta_N(E)d[E]\,.
\end{equation}
We pursue an idea similar to the one by Basor and Forrester \cite{BasFor94}. We supplement the factor
\begin{equation}\label{1.3}
 \displaystyle \Delta_N(E)\prod\limits_{a=1}^N\prod\limits_{b=1}^{k}\frac{E_a-\kappa_{b2}}{E_a-\kappa_{b1}}
\end{equation}
by 
\begin{equation}\label{1.4}
 \displaystyle \sqrt{\Ber^{(2)}_{k/k}(\kappa)}=\frac{\Delta_{k}(\kappa_1)\Delta_{k}(\kappa_2)}{\prod\limits_{a,b=1}^{k}(\kappa_{a1}-\kappa_{b2})}\,.
\end{equation}
Both factors together are up to a sign
\begin{equation}\label{1.5}
 \displaystyle \sqrt{\Ber^{(2)}_{k/k+N}(\kappa_{1};\kappa_{2},E)}=\pm\frac{\Delta_{k}(\kappa_1)\Delta_{k+N}(\kappa_2,E)}{\prod\limits_{a,b=1}^{k}(\kappa_{a1}-\kappa_{b2})\prod\limits_{a=1}^k\prod\limits_{b=1}^{N}(\kappa_{a1}-E_b)}\,.
\end{equation}
The authors in Ref. \cite{BasFor94} have shown that this function has for all $N\in\mathbb{N}_0$ a determinantal structure mixing terms of the Vandermonde determinant and the Cauchy determinant,
\begin{equation}\label{1.6}
 \displaystyle \sqrt{\Ber^{(2)}_{k/k+N}(\kappa_{1};\kappa_{2},E)}=\pm\det\left[\begin{array}{c|c} \displaystyle\frac{1}{\kappa_{a1}-\kappa_{b2}} & \displaystyle\underset{}{\frac{1}{\kappa_{a1}-E_{b}}} \\ \hline \displaystyle\overset{}{\kappa_{b2}^{a-1}} & \displaystyle E_b^{a-1} \end{array} \right]\,.
\end{equation}
The insight crucial for this work and not contained in Ref. \cite{BasFor94} is the intimate connection of Eq.~\eref{1.6} to superspace: $\Ber^{(2)}_{p/q}$ is the Jacobian or Berezinian for the diagonalization of symmetric matrices or supermatrices, respectively, as shown in Refs. \cite{Guh91}. We will prove Eq.~\eref{1.6} in a new way and obtain also an interesting intermediate result not given in Ref. \cite{BasFor94}, see Sec. \ref{sec2}.

We now proceed with the evaluation of $Z(\kappa)$. We shift the eigenvalue integrals into the determinant and obtain
\begin{equation}\label{1.7}
 \displaystyle Z(\kappa)=\frac{c}{\sqrt{\Ber^{(2)}_{k/k}(\kappa)}}\det\left[\begin{array}{c|c} \displaystyle\frac{1}{\kappa_{a1}-\kappa_{b2}} & \displaystyle\underset{}{F_b(\kappa_{a1})} \\ \hline \displaystyle\overset{}{\kappa_{b2}^{a-1}} & \displaystyle M_{ab} \end{array} \right]\,.
\end{equation}
The symmetric matrix $M_{ab}$ comprises the moments of the probability density $P$ and the functions $F_b$ are the Cauchy transform of those moments. At this point we have a choice for how to proceed further. For instance we can reorder the monomials in the entries of the determinant to orthogonal polynomials with respect to $P$. Then, $M_{ab}$ becomes diagonal and $F_b$ are the Cauchy transforms of the orthogonal polynomials. Thus we arrive at the well known result, see Ref.~\cite{BDS03}. On the other hand, we can choose an arbitrary set of linearly independent polynomials. Then, we use the important property of the determinant
\begin{equation}\label{1.7b}
 \det\left[\begin{array}{cc} A & B \\ C & D \end{array}\right]=\det D \det[A-BD^{-1}C]
\end{equation}
for arbitrary matrices $A$, $B$ and $C$ and an invertible quadratic matrix $D$. This finally yields
\begin{eqnarray}
 Z(\kappa)&=&\displaystyle \frac{c}{\sqrt{\Ber^{(2)}_{k/k}(\kappa)}}\det\left[\frac{1}{\kappa_{a1}-\kappa_{b2}} -\sum\limits_{m,n=1}^N F_m(\kappa_{a1})M_{nm}^{-1}\kappa_{b2}^{n-1}\right]=\nonumber\\
 &=&\displaystyle \frac{c}{\sqrt{\Ber^{(2)}_{k/k}(\kappa)}}\det\,K(\kappa_{a1},\kappa_{b2})\,.\label{1.8}
\end{eqnarray}
We obtain the correct result \cite{TraWid98} without the Mehta-Mahoux theorem for an arbitrary choice of polynomials. The orthogonal polynomials are not the tool to identify the determinantal structures. They are a result of the calculation.

In the next sections, we extend this sketch to a careful discussion for a large class of integrals. We will see that determinantal structures derived in many different fields of random matrix theory have a common origin.

\section{Determinantal structure of Berezinians}\label{sec2}

In Sec. \ref{sec2.1}, we investigate the determinantal structure of the Berezinians resulting from Hermitian supermatrices. These Berezinians are crucial for the calculations in the ensuing sections. For the sake of completeness, we give the determinantal structure according to the supergroup $\UOSp(p/q)$ in Sec. \ref{sec2.2}.

\subsection{Berezinians related to the supergroup $\U(p/q)$}\label{sec2.1}

As we have seen in Sec.~\ref{sec1.b}, Berezinians resulting from diagonalization of supermatrices play a role of paramount importance for our method. Although we do not use any integral in superspace we find those Berezinians in the ratios of characteristic polynomials times the Vandermonde determinant. The crucial step is here to understand that those Berezinians have always a determinantal structure.

For the Vandermonde determinant the determinantal structure
\begin{equation}
 \Delta_k(\kappa)=\displaystyle\prod\limits_{1\leq a<b\leq k}\left(\kappa_{a}-\kappa_{b}\right)=(-1)^{k(k-1)/2}\det\left[\kappa_{b}^{a-1}\right]_{1\leq a,b\leq k}\label{2.1}\,,
\end{equation}
has been known for a long time \cite{Meh67}. Moreover, the square root of the Berezinian resulting from a diagonalization of the supersymmetric analog of a $(k+k)\times(k+k)$ Hermitian matrix is known \cite{Guh91}, up to a sign, to be equal to the Cauchy determinant
 \begin{eqnarray}
  \fl&&\displaystyle\sqrt{\Ber^{(2)}_{k/k}(\kappa)}=\displaystyle\frac{\prod\limits_{1\leq a<b\leq k}\left(\kappa_{a1}-\kappa_{b1}\right)\left(\kappa_{a2}-\kappa_{b2}\right)}{\prod\limits_{1\leq a,b\leq k}\left(\kappa_{a1}-\kappa_{b2}\right)}=\displaystyle(-1)^{k(k-1)/2}\det\left[\frac{1}{\kappa_{a1}-\kappa_{b2}}\right]_{1\leq a,b\leq k}\,.\nonumber\\
  \fl\label{2.3}
\end{eqnarray}
The last equality sign is due to Cauchy's lemma \cite{PolSze25}. The upper index ``$(2)$'' refers to the Dyson index $\beta\in\{1,2,4\}$ and indicates that this Berezinian is related to Hermitian supermatrices.

The next step is to generalize these structures to an arbitrary number of bosonic eigenvalues $\kappa_1=\diag(\kappa_{11},\ldots,\kappa_{p1})$ and of fermionic eigenvalues $\kappa_2=\diag(\kappa_{12},\ldots,\kappa_{q2})$. In \ref{app1.1}, we derive the determinantal structures of the Berezinians, see Ref.~\cite{Guh96},
 \begin{equation}
  \displaystyle\sqrt{{\rm Ber}_{p/q}^{(2)}(\kappa)}=\displaystyle\frac{\prod\limits_{1\leq a<b\leq p}\left(\kappa_{a1}-\kappa_{b1}\right)\prod\limits_{1\leq a<b\leq q}\left(\kappa_{a2}-\kappa_{b2}\right)}{\prod\limits_{a=1}^p\prod\limits_{b=1}^q\left(\kappa_{a1}-\kappa_{b2}\right)}\label{2.5}\,,
 \end{equation}
for arbitrary $p$ and $q$. Equation~\eref{2.3} is the case $p=q=k$. Under the condition $p\leq q$, we obtain
 \begin{equation}
  \displaystyle\sqrt{{\rm Ber}_{p/q}^{(2)}(\kappa)}=
  (-1)^{q(q-1)/2+(q+1)p}\det\left[\begin{array}{c}\left\{\displaystyle\frac{\kappa_{a1}^{p-q}\kappa_{b2}^{q-p}}{\kappa_{a1}-\kappa_{b2}}\right\}\underset{1\leq b\leq q}{\underset{1\leq a\leq p}{\ }} \\ \left\{\kappa_{b2}^{a-1}\right\}\underset{1\leq b\leq q}{\underset{1\leq a\leq q-p}{\ }} \end{array}\right]\label{2.7}\,.
 \end{equation}
Since the left hand side is up to $(-1)^{pq}$ symmetric under exchanging  the bosonic eigenvalues with the fermionic ones, the condition $p\leq q$ is not a restriction. This result is similar to Eq.~\eref{1.6}. In Sec.~\ref{sec5.2} and \ref{app5},  we show  that it is useful for some calculations.

The left hand side of Eq.~\eref{2.7} is translation invariant $\kappa\to\kappa+\varepsilon\eins_{p+q}$ with a constant $\varepsilon$. Thus, we may shift the expressions on the right hand side by $\varepsilon$,
 \begin{eqnarray}
  \fl\displaystyle\sqrt{{\rm Ber}_{p/q}^{(2)}(\kappa)}&=&
  (-1)^{q(q-1)/2+(q+1)p}\det\left[\begin{array}{c}\left\{\displaystyle\frac{(\kappa_{a1}+\varepsilon)^{p-q}(\kappa_{b2}+\varepsilon)^{q-p}}{\kappa_{a1}-\kappa_{b2}}\right\}\underset{1\leq b\leq q}{\underset{1\leq a\leq p}{\ }} \\ \left\{(\kappa_{b2}+\varepsilon)^{a-1}\right\}\underset{1\leq b\leq q}{\underset{1\leq a\leq q-p}{\ }} \end{array}\right]\,.\label{2.10}
 \end{eqnarray}
We expand the entries in the lower $(q-p)\times q$ block in $\varepsilon$. We notice that all rows together are a linearly independent set of polynomials from order zero to order $q-p-1$. As the determinant is skew symmetric, it yields
 \begin{eqnarray}
  \fl\displaystyle\sqrt{{\rm Ber}_{p/q}^{(2)}(\kappa)}&=&
  (-1)^{q(q-1)/2+(q+1)p}\det\left[\begin{array}{c}\left\{\displaystyle\frac{(\kappa_{a1}+\varepsilon)^{p-q}(\kappa_{b2}+\varepsilon)^{q-p}}{\kappa_{a1}-\kappa_{b2}}\right\}\underset{1\leq b\leq q}{\underset{1\leq a\leq p}{\ }} \\ \left\{\kappa_{b2}^{a-1}\right\}\underset{1\leq b\leq q}{\underset{1\leq a\leq q-p}{\ }} \end{array}\right]\label{2.11}\,.
 \end{eqnarray}
Since $\varepsilon$ is arbitrary we take the limit for $\varepsilon$ to infinity and obtain the final result
 \begin{eqnarray}
  \fl\displaystyle\sqrt{{\rm Ber}_{p/q}^{(2)}(\kappa)}&=&
  (-1)^{q(q-1)/2+(q+1)p}\det\left[\begin{array}{c}\left\{\displaystyle\frac{1}{\kappa_{a1}-\kappa_{b2}}\right\}\underset{1\leq b\leq q}{\underset{1\leq a\leq p}{\ }} \\ \left\{\kappa_{b2}^{a-1}\right\}\underset{1\leq b\leq q}{\underset{1\leq a\leq q-p}{\ }} \end{array}\right]\label{2.12}
 \end{eqnarray}
which is identical to the result of Basor and Forrester \cite{BasFor94}. Indeed, Eq.~\eref{2.12} does not exhibit the nice symmetry between the bosonic and fermionic eigenvalues as in Eqs.~\eref{2.3} and \eref{2.5}.

\subsection{Berezinians related to the supergroup $\UOSp(p/q)$}\label{sec2.2}

As for the Vandermonde determinant itself, the determinantal structure for the forth power thereof is also well known \cite{Meh67},
\begin{eqnarray}
 \Delta_k^4(\kappa)&=& \displaystyle\prod\limits_{1\leq a<b\leq k}\left(\kappa_{a2}-\kappa_{b2}\right)^4=\nonumber\\
 &=&\displaystyle \det\left[\begin{array}{c|c} \kappa_{b}^{a-1} & (a-1)\kappa_{b}^{a-2} \end{array}\right]\underset{1\leq b\leq k}{\underset{1\leq a\leq 2k}{\ }}\label{2.2}\,. 
\end{eqnarray}
Recently \cite{KGG08}, it was shown that the Berezinian corresponding to the supergroup $\UOSp(2k/k)$ has a determinantal structure, too,
 \begin{eqnarray}
  \fl{\rm Ber}_{2k/k}^{(1)}(\kappa)&=&{\rm Ber}_{k/2k}^{(4)}(\tilde{\kappa})=\displaystyle\displaystyle\frac{\prod\limits_{1\leq a<b\leq 2k}\left(\kappa_{a1}-\kappa_{b1}\right)\prod\limits_{1\leq a<b\leq k}\left(\kappa_{a2}-\kappa_{b2}\right)^4}{\prod\limits_{a=1}^{2k}\prod\limits_{b=1}^k\left(\kappa_{a1}-\kappa_{b2}\right)^2}=\nonumber\\
   \fl&=& \displaystyle \det\left[\begin{array}{c|c} \displaystyle\frac{1}{\kappa_{a1}-\kappa_{b2}} & \displaystyle\frac{1}{(\kappa_{a1}-\kappa_{b2})^2} \end{array}\right]\underset{1\leq b\leq k}{\underset{1\leq a\leq 2k}{\ }}\label{2.4}\,.
 \end{eqnarray}
Here, we define $\tilde{\kappa}=\diag(\kappa_2,\kappa_1)$.

In \ref{app1.2}, we derive the analog of Eq.~\eref{2.5} for the Berezinian
\begin{eqnarray}
  \fl{\rm Ber}_{p/q}^{(1)}(\kappa)&=&{\rm Ber}_{q/p}^{(4)}(\tilde{\kappa})=\displaystyle\displaystyle\frac{\prod\limits_{1\leq a<b\leq p}\left(\kappa_{a1}-\kappa_{b1}\right)\prod\limits_{1\leq a<b\leq q}\left(\kappa_{a2}-\kappa_{b2}\right)^4}{\prod\limits_{a=1}^{p}\prod\limits_{b=1}^q\left(\kappa_{a1}-\kappa_{b2}\right)^2}\label{2.6}\,,
\end{eqnarray}
see Ref.~\cite{KohGuh05}. Here, we have to distinguish between $p\leq2q$ and $p\geq2q$. For the first case, we obtain
 \begin{eqnarray}
  \fl{\rm Ber}_{p/q}^{(1)}(\kappa)&=&{\rm Ber}_{q/p}^{(4)}(\tilde{\kappa})=\nonumber\\
  \fl&=& \displaystyle(-1)^{p} \det\left[\begin{array}{c|c}\left\{ \displaystyle\frac{\kappa_{a1}^{p-2q}\kappa_{b2}^{2q-p}}{\kappa_{a1}-\kappa_{b2}}\right. & \left.\displaystyle\frac{\partial}{\partial\kappa_{b2}}\frac{\kappa_{a1}^{p-2q}\kappa_{b2}^{2q-p}}{\kappa_{a1}-\kappa_{b2}} \right\}\underset{1\leq b\leq q}{\underset{1\leq a\leq p}{\ }} \\ \left\{ \kappa_{b2}^{a-1}\right. & \left.(a-1)\kappa_{b2}^{a-2} \right\}\underset{1\leq b\leq q}{\underset{1\leq a\leq 2q-p}{\ }} \end{array}\right]\label{2.8}
 \end{eqnarray}
and, for the second one, we get
 \begin{eqnarray}
  \fl{\rm Ber}_{p/q}^{(1)}(\kappa)&=&{\rm Ber}_{q/p}^{(4)}(\tilde{\kappa})=\label{2.9}\\
   \fl&=&  \displaystyle(-1)^{p(p-1)/2+q}\det\left[\begin{array}{c|cc}\left\{\displaystyle\frac{\kappa_{a1}^{p-2q}\kappa_{b2}^{2q-p}}{\kappa_{a1}-\kappa_{b2}}\right. & \left.\displaystyle\frac{\kappa_{a1}^{p-2q}\kappa_{b2}^{2q-p}}{(\kappa_{a1}-\kappa_{b2})^2}\right\}\underset{1\leq b\leq q}{\underset{1\leq a\leq p}{\ }} & \left\{\kappa_{a1}^{b-1}\right\}\underset{1\leq b\leq p-2q}{\underset{1\leq a\leq p}{\ }} \end{array}\right]\hspace*{-0.1cm}.\nonumber
 \end{eqnarray}
We apply the same procedure as in Sec.~\ref{sec2.1} and shift all elements by $\varepsilon$. Taking the limit $\varepsilon\to\infty$, we find
 \begin{eqnarray}
  \fl{\rm Ber}_{p/q}^{(1)}(\kappa)&=&{\rm Ber}_{q/p}^{(4)}(\tilde{\kappa})=\nonumber\\
  \fl&=& \displaystyle(-1)^{p} \det\left[\begin{array}{c|c}\left\{ \displaystyle\frac{1}{\kappa_{a1}-\kappa_{b2}}\right. & \left.\displaystyle\frac{1}{(\kappa_{a1}-\kappa_{b2})^2} \right\}\underset{1\leq b\leq q}{\underset{1\leq a\leq p}{\ }} \\ \left\{ \kappa_{b2}^{a-1}\right. & \left.(a-1)\kappa_{b2}^{a-2} \right\}\underset{1\leq b\leq q}{\underset{1\leq a\leq 2q-p}{\ }} \end{array}\right]\label{2.13}
 \end{eqnarray}
for $p\leq2q$ and
 \begin{eqnarray}
  \fl{\rm Ber}_{p/q}^{(1)}(\kappa)&=&{\rm Ber}_{q/p}^{(4)}(\tilde{\kappa})=\label{2.14}\\
   \fl&=&  \displaystyle(-1)^{p(p-1)/2+q}\det\left[\begin{array}{c|cc}\left\{\displaystyle\frac{1}{\kappa_{a1}-\kappa_{b2}}\right. & \left.\displaystyle\frac{1}{(\kappa_{a1}-\kappa_{b2})^2}\right\}\underset{1\leq b\leq q}{\underset{1\leq a\leq p}{\ }} & \left\{\kappa_{a1}^{b-1}\right\}\underset{1\leq b\leq p-2q}{\underset{1\leq a\leq p}{\ }} \end{array}\right]\hspace*{-0.1cm}.\nonumber
 \end{eqnarray}
for $p\geq2q$. We notice that the determinantal structure of the ordinary Jacobians which are powers of Vandermonde determinants mixes with the structure of the Cauchy determinant, as for the $\U(p/q)$ case.

\section{Main result}\label{sec3}

We discuss two types of integrals which cover averages over ratios of characteristic polynomials for a large class of matrix ensembles. For both types we find determinantal structures. These types are integrals with a square root of a Berezinian \eref{2.5} and with a Vandermonde determinant to the second power. They are studied in Secs.~\ref{sec3.1} and \ref{sec3.2}, respectively.

\subsection{Integrals of square root--Berezinian type}\label{sec3.1}

We consider the integral
\begin{eqnarray}
 \fl Z_{k_1/k_2}^{(N_1/N_2)}(\kappa)&=&\int\limits_{\mathbb{C}^{N_1+N_2}}\prod\limits_{j=1}^{N_1}g_j(z_{j1})\prod\limits_{j=1}^{N_2}f_j(z_{j2})\times\nonumber\\
 \fl &\times&\frac{\prod\limits_{a=1}^{N_1}\prod\limits_{b=1}^{k_1}(z_{a1}-\kappa_{b1})\prod\limits_{a=1}^{N_2}\prod\limits_{b=1}^{k_2}(z_{a2}-\kappa_{b2})}{\prod\limits_{a=1}^{N_1}\prod\limits_{b=1}^{k_2}(z_{a1}-\kappa_{b2})\prod\limits_{a=1}^{k_1}\prod\limits_{b=1}^{N_2}(\kappa_{a1}-z_{b2})}\sqrt{\Ber_{N_1/N_2}^{(2)}(z)}d[z]\,,\label{3.1}
\end{eqnarray}
where the $z_j$ are complex variables. The functions $g_j$ and $f_j$ and the variables $\kappa$ are chosen in such a way that the integral is convergent. The measure $d[z]$ is the product of the differentials of the real and imaginary parts. The applications which we will give are particular cases of this integral, although these applications correspond to essentially different ensembles. Thus, we show a fundamental relation which yields determinantal structures.

The crucial step is to extend the integral \eref{3.1} by $\sqrt{\Ber_{k_1/k_2}^{(2)}(\kappa)}$ and to recognize that we obtain a new Berezinian
\begin{eqnarray}
  \fl Z_{k_1/k_2}^{(N_1/N_2)}(\kappa)&=&\int\limits_{\mathbb{C}^{N_1+N_2}}\prod\limits_{j=1}^{N_1}g_j(z_{j1})\prod\limits_{j=1}^{N_2}f_j(z_{j2})\frac{\sqrt{\Ber_{N_1+k_1/N_2+k_2}^{(2)}(\tilde{z})}}{\sqrt{\Ber_{k_1/k_2}^{(2)}(\kappa)}}d[z]\,,\label{3.2}
\end{eqnarray}
where the new bosonic eigenvalues are $\tilde{z}_1=\diag(z_1,\kappa_1)$ and the new fermionic eigenvalues are $\tilde{z}_2=\diag(z_2,\kappa_2)$. Now we use the determinantal structure of the square root Berezinian shown in Sec.~\ref{sec2.1}.

Without loss of generality, we assume $N_2\geq N_1$. In \ref{app2.2}, we show the details of this calculation and only give the results here. The simplest case  is $k_1=k_2=k$. Then, the condition $(N_1+k_1)\leq (N_2+k_2)$ is automatically fulfilled. The integral \eref{3.2} is then a quotient of two determinants times a constant
\begin{eqnarray}
  \fl Z_{k/k}^{(N_1/N_2)}(\kappa)&=&\frac{(-1)^{(N_2+k)(N_2+k-1)/2}\det\mathbf{M}_{N_1/N_2}}{\sqrt{\Ber_{k/k}^{(2)}(\kappa)}}\det\left[K^{(N_1/N_2)}(\kappa_{a1},\kappa_{b2})\right]_{1\leq a,b\leq k}\,,\label{3.3}
\end{eqnarray}
where we define
\begin{eqnarray}
   \mathbf{G}^{(N_1/N_2)}(\kappa_{b2})&=&\left[\begin{array}{c} \left\{\kappa_{b2}^{a-1}\right\}\underset{1\leq a\leq N_2-N_1}{\ } \\ \displaystyle\left\{\int\limits_{\mathbb{C}}\frac{g_a(z)}{z-\kappa_{b2}}d[z]\right\}\underset{1\leq a\leq N_1}{\ } \end{array}\right]\label{3.5}\,,\\
   \mathbf{F}^{(N_2)}(\kappa_{a1})&=&\left[\int\limits_{\mathbb{C}}\frac{f_b(z)}{\kappa_{a1}-z}d[z]\right]\underset{1\leq b\leq N_2}{\ }\label{3.7}\,,\\
   \mathbf{M}_{N_1/N_2}&=&\left[\begin{array}{c}  \left\{\int\limits_{\mathbb{C}}f_b(z)z^{a-1}d[z]\right\}\underset{1\leq b\leq N_2}{\underset{1\leq a\leq N_2-N_1}{\ }} \\  \displaystyle\left\{\int\limits_{\mathbb{C}^2}\frac{g_a(z_1)f_b(z_2)}{z_1-z_2}d[z]\right\}\underset{1\leq b\leq N_2}{\underset{1\leq a\leq N_1}{\ }}  \end{array}\right]\label{3.10}\,,\\
   K^{(N_1/N_2)}(\kappa_{a1},\kappa_{b2})&=&\frac{1}{\kappa_{a1}-\kappa_{a2}}-\mathbf{F}^{(N_2)}(\kappa_{a1})\mathbf{M}_{N_1/N_2}^{-1}\mathbf{G}^{(N_1/N_2)}(\kappa_{b2})\label{3.11}\,.
\end{eqnarray}
Since the entries $K^{(N_1/N_2)}(\kappa_{a1},\kappa_{b2})$ are independent of the dimension $k$, we identify 
\begin{equation}\label{3.12}
 K^{(N_1/N_2)}(\kappa_{a1},\kappa_{b2})=\frac{(-1)^{N_2(N_2+1)/2}}{\det\mathbf{M}_{N_1/N_2}}\frac{Z_{1/1}^{(N_1/N_2)}(\kappa_{a1},\kappa_{b2})}{\kappa_{a1}-\kappa_{b2}}
\end{equation}
which is the case $k=1$. The normalization constant follows from $k=0$ and is given by
\begin{equation}\label{3.13}
 C_{N_1/N_2}=Z_{0/0}^{(N_1/N_2)}=(-1)^{N_2(N_2-1)/2}\det\mathbf{M}_{N_1/N_2}\,.
\end{equation}
 This leads to the very compact result
\begin{eqnarray}
  \fl&& Z_{k/k}^{(N_1/N_2)}(\kappa)=\frac{(-1)^{k(k-1)/2}}{C_{N_1/N_2}^{k-1}\sqrt{\Ber_{k/k}^{(2)}(\kappa)}}\det\left[\frac{Z_{1/1}^{(N_1/N_2)}(\kappa_{a1},\kappa_{b2})}{\kappa_{a1}-\kappa_{b2}}\right]_{1\leq a,b\leq k}\,.\label{3.14}
\end{eqnarray}
We recall that the functions $g_j$ and $f_j$ are arbitrary. This means that the fundamental structure of the ratios of the characteristic polynomials times the Berezinian fully generates the whole determinantal expression.

The cases $k_1\leq k_2$ and $k_1\geq k_2$ cover all cases mentioned above. As in Ref.~\cite{SWG99}, we trace these cases back by introducing $|k_1-k_2|$ dummy variables. These variables enlarge the $(k_1+k_2)\times(k_1+k_2)$ eigenvalue matrix of $\kappa$ to a $(k+k)\times(k+k)$ eigenvalue matrix, where $k=\max\{k_1,k_2\}$. Then, we use our result obtained above and remove these additional eigenvalues. The explicit calculations are given in \ref{app2.2} and \ref{app2.3}. We obtain
\begin{eqnarray}
 \fl&&\displaystyle Z_{k_1/k_2}^{(N_1/N_2)}(\kappa)=\frac{(-1)^{k_1(k_1-1)/2+(k_2-k_1)N_1}}{C_{N_1/N_2}^{k_2-1}\sqrt{\Ber_{k_1/k_2}^{(2)}(\kappa)}}\times\nonumber\\
 \fl&\times&\det\left[\begin{array}{c} \left\{\displaystyle\frac{Z_{1/1}^{(N_1/N_2)}(\kappa_{a1},\kappa_{b2})}{\kappa_{a1}-\kappa_{b2}}\right\}\underset{1\leq b\leq k_2}{\underset{1\leq a\leq k_1}{\ }} \\ \left\{\displaystyle\left.\left(\kappa_0^2\frac{\partial}{\partial \kappa_0}\right)^{a-1}\frac{\kappa_{0}^{N_2-N_1+1}Z_{1/1}^{(N_1/N_2)}(\kappa_{0},\kappa_{b2})}{\kappa_{0}-\kappa_{b2}}\right|_{\kappa_0\to\infty}\right\}\underset{1\leq b\leq k_2}{\underset{1\leq a\leq k_2-k_1}{\ }} \end{array}\right]\label{3.15}
\end{eqnarray}
for $k_1\leq k_2$ and
\begin{eqnarray}
 \fl&&\displaystyle Z_{k_1/k_2}^{(N_1/N_2)}(\kappa)=\frac{(-1)^{(k_2+2k_1)(k_2-1)/2+(k_1-k_2)(N_2-N_1)}}{C_{N_1/N_2}^{k_1-1}\sqrt{\Ber_{k_1/k_2}^{(2)}(\kappa)}}\times\nonumber\\
 \fl&\times&\det\left[\begin{array}{c} \left\{\displaystyle\frac{Z_{1/1}^{(N_1/N_2)}(\kappa_{b1},\kappa_{a2})}{\kappa_{b1}-\kappa_{a2}}\right\}\underset{1\leq b\leq k_1}{\underset{1\leq a\leq k_2}{\ }} \\ \left\{\displaystyle\left.\left(\kappa_0^2\frac{\partial}{\partial \kappa_0}\right)^{a-1}\frac{\kappa_{0}^{N_1-N_2+1}Z_{1/1}^{(N_1/N_2)}(\kappa_{b1},\kappa_{0})}{\kappa_{b1}-\kappa_{0}}\right|_{\kappa_0\to\infty}\right\}\underset{1\leq b\leq k_1}{\underset{1\leq a\leq k_1-k_2}{\ }} \end{array}\right]\label{3.16}
\end{eqnarray}
for $k_1\geq k_2$. For $|k_1-k_2|=1$ the average in the last row is only over one characteristic polynomial, i.e. it is equal to $Z_{0/1}^{(N_1/N_2)}(\kappa_{b2})$ in Eq.~\eref{3.15} and $Z_{1/0}^{(N_1/N_2)}(\kappa_{b1})$ in Eq.~\eref{3.16}. The limits in Eqs.~\eref{3.15} and \eref{3.16} are well defined, as a comparison with Eq.~\eref{3.1} shows, and can be calculated by writing the derivative as a contour integral around $1/\kappa_0=0$. The limits are explicitly given as
\begin{eqnarray}
 \fl\displaystyle&&\left.\left(\kappa_0^2\frac{\partial}{\partial \kappa_0}\right)^{a-1}\frac{\kappa_{0}^{N_2-N_1+1}Z_{1/1}^{(N_1/N_2)}(\kappa_{0},\kappa_{b2})}{\kappa_{0}-\kappa_{b2}}\right|_{\kappa_0\to\infty}=\nonumber\\
 \fl&=&(-1)^{a-1+N_2}(a-1)!C_{N_1/N_2}\left[\kappa_{b2}^{a-1+N_2-N_1}-\mathbf{f}_a\mathbf{M}_{N_1/N_2}^{-1}\mathbf{G}^{(N_1/N_2)}(\kappa_{b2})\right]\label{3.16a}
\end{eqnarray}
and
\begin{eqnarray}
 \fl\displaystyle&&\left.\left(\kappa_0^2\frac{\partial}{\partial \kappa_0}\right)^{a-1}\frac{\kappa_{0}^{N_1-N_2+1}Z_{1/1}^{(N_1/N_2)}(\kappa_{b1},\kappa_{0})}{\kappa_{b1}-\kappa_{0}}\right|_{\kappa_0\to\infty}=(-1)^{a+N_2}(a-1)!C_{N_1/N_2}\times\nonumber\\
 \fl&\times&\left[\kappa_{b1}^{a-1+N_1-N_2}\Theta(a+N_1-N_2-1)-\mathbf{F}^{(N_2)}(\kappa_{b1})\mathbf{M}_{N_1/N_2}^{-1}\mathbf{g}_{a}\right]\label{3.16b}\,,
\end{eqnarray}
where we define the matrices
\begin{eqnarray}
 \mathbf{f}_{a}&=&\left[\int\limits_{\mathbb{C}}f_b(z)z^{a-1+N_2-N_1}d[z]\right]\underset{1\leq b\leq N_2}{\ }\label{3.16c}\,,\\
 \mathbf{g}_{a}&=&\left[\begin{array}{c} \left\{-\delta_{N_2-N_1+1-a,b}\right\}\underset{1\leq b\leq N_2-N_1}{\ } \\ \displaystyle\left\{\int\limits_{\mathbb{C}}g_b(z)z^{a-1+N_1-N_2}d[z]\Theta(a+N_1-N_2-1)\right\}\underset{1\leq b\leq N_1}{\ } \end{array}\right]\label{3.16d}\,.
\end{eqnarray}
The function $\Theta$ is the Heaviside distribution for discrete numbers which means it is the integrated Kronecker-$\delta$ and, hence, unity at zero.

\subsection{Integrals of squared--Vandermonde type}\label{sec3.2}

Now, we investigate integrals of the type
\begin{equation}
 \fl\displaystyle \widetilde{Z}_{\frac{k_1/k_2}{\overset{}{l_1/l_2}}}^{(N)}(\kappa,\lambda)=\int\limits_{\mathbb{C}^{N}}\prod\limits_{j=1}^{N}g(z_{j}) \frac{\prod\limits_{a=1}^{k_2}\prod\limits_{b=1}^{N}(\kappa_{a2}-z_{b})\prod\limits_{a=1}^{l_2}\prod\limits_{b=1}^{N}(\lambda_{a2}-z_{b}^*)}{\prod\limits_{a=1}^{k_1}\prod\limits_{b=1}^{N}(\kappa_{a1}-z_{b})\prod\limits_{a=1}^{l_1}\prod\limits_{b=1}^{N}(\lambda_{a1}-z_{b}^*)}|\Delta_N(z)|^2d[z]\,.\label{3.17}
\end{equation}
The function $g$ is, as in Sec. \ref{sec3.1}, an arbitrary function with the only restriction that the integral above is convergent. Instead of one eigenvalue set as in the subsection above, we have now two eigenvalue sets $\kappa=\diag(\kappa_{11},\ldots,\kappa_{k_11},\kappa_{12},\ldots,\kappa_{k_22})$ and $\lambda=\diag(\lambda_{11},\ldots,\lambda_{l_11},\lambda_{12},\ldots,\lambda_{l_22})$. Because of these two sets, we have to extend the fraction by two square roots of Berezinians and find
\begin{eqnarray}
 \fl\widetilde{Z}_{\frac{k_1/k_2}{\overset{}{l_1/l_2}}}^{(N)}(\kappa,\lambda) &=&\int\limits_{\mathbb{C}^{N}}\prod\limits_{j=1}^{N}g(z_{j}) \frac{\sqrt{\Ber_{k_1/k_2+N}^{(2)}(\tilde{z})}\sqrt{\Ber_{l_1/l_2+N}^{(2)}(\hat{z})}}{\sqrt{\Ber_{k_1/k_2}^{(2)}(\kappa)}\sqrt{\Ber_{l_1/l_2}^{(2)}(\lambda)}}d[z]\,.\label{3.18}
\end{eqnarray}
We introduce $\tilde{z}=\diag(\kappa_1,\kappa_2,z)$ and $\hat{z}=\diag(\lambda_1,\lambda_2,z^*)$.

To integrate Eq. \eref{3.18}, we first discuss the case $d=k_2+N-k_1=l_2+N-l_1\geq 0$. Under the integral we have two determinants with $N$ rows depending on one $z_a$ or one $z_a^*$. The other rows are independent of any $z_a$ and $z_a^*$. Thus, we use an integration theorem similar to Andr\'{e}ief's \cite{And1883} which we derive in \ref{app3.1}. In \ref{app4} we carry out the calculation and find
\begin{eqnarray}
 \fl&&\widetilde{Z}_{\frac{k_1/k_2}{\overset{}{l_1/l_2}}}^{(N)}(\kappa,\lambda)=\frac{(-1)^{(l_2+k_2)(l_1+k_1-1)/2}N!\det\mathbf{\widetilde{M}}_{d}}{\sqrt{\Ber_{k_1/k_2}^{(2)}(\kappa)}\sqrt{\Ber_{l_1/l_2}^{(2)}(\lambda)}} \times\nonumber\\
 \fl&\times&\det\left[\begin{array}{cc} \left\{\widetilde{K}_{11}^{(d)}(\lambda_{a2},\kappa_{b2})\right\}\underset{1\leq b\leq k_2}{\underset{1\leq a\leq l_2}{\ }} & \left\{\widetilde{K}_{12}^{(d)}(\lambda_{b1},\lambda_{a2})\right\}\underset{1\leq b\leq l_1}{\underset{1\leq a\leq l_2}{\ }} \\  \left\{\widetilde{K}_{21}^{(d)}(\kappa_{a1},\kappa_{b2})\right\}\underset{1\leq b\leq k_2}{\underset{1\leq a\leq k_1}{\ }} & \left\{\widetilde{K}_{22}^{(d)}(\kappa_{a1},\lambda_{b1})\right\}\underset{1\leq b\leq l_1}{\underset{1\leq a\leq k_1}{\ }}  \end{array}\right]\,,\label{3.19}
\end{eqnarray}
where
\begin{eqnarray}
   \widetilde{Z}_{\frac{1/0}{\overset{}{1/0}}}^{(1)}(\kappa_{a1},\lambda_{b1})&=&\int\limits_{\mathbb{C}}\frac{g(z)}{(\kappa_{a1}-z)(\lambda_{b1}-z^*)}d^2z\,,\label{3.20}\\
   \mathbf{\widetilde{F}}_{d}(\kappa_{a1})&=&\left[\int\limits_{\mathbb{C}}\frac{g(z)z^{*\,b-1}}{\kappa_{a1}-z}d^2z\right]_{1\leq b\leq d}\label{3.21}\,,\\
   \mathbf{\widetilde{F}^{\,(*)}}_{d}(\lambda_{b1})&=&\left[\int\limits_{\mathbb{C}}\frac{g(z)z^{a-1}}{\lambda_{b1}-z^*}d^2z\right]_{1\leq a\leq d}\label{3.22}\,,\\
   \mathbf{\widetilde{M}}_{d}&=&\left[\int\limits_{\mathbb{C}}g(z)z^{a-1}z^{*\,b-1}d^2z\right]_{1\leq a,b\leq d}\,,\label{3.23}\\
   \mathbf{\Lambda}_{d}(\lambda_{a2})&=&\left[\lambda_{a2}^{b-1}\right]_{1\leq b\leq d}\,,\label{3.24}\\
   \mathbf{K}_{d}(\kappa_{b2})&=&\left[\kappa_{b2}^{a-1}\right]_{1\leq a\leq d}\,,\label{3.25}\\
   \widetilde{K}_{11}^{(d)}(\kappa_{b2},\lambda_{a2})&=&-\mathbf{\Lambda}_{d}(\lambda_{a2})\mathbf{\widetilde{M}}_{d}^{-1}\mathbf{K}_{d}(\kappa_{b2})\,,\label{3.26}\\
   \widetilde{K}_{12}^{(d)}(\lambda_{b1},\lambda_{a2})&=&\displaystyle\frac{1}{\lambda_{b1}-\lambda_{a2}}-\mathbf{\Lambda}_{d}(\lambda_{a2})\mathbf{\widetilde{M}}_{d}^{-1}\mathbf{\widetilde{F}^{\,(*)}}_{d}(\lambda_{b1})\,,\label{3.27}\\
   \widetilde{K}_{21}^{(d)}(\kappa_{a1},\kappa_{b2})&=&\displaystyle\frac{1}{\kappa_{a1}-\kappa_{b2}}-\mathbf{\widetilde{F}}_{d}(\kappa_{a1})\mathbf{\widetilde{M}}_{d}^{-1}\mathbf{K}_{d}(\kappa_{b2})\label{3.28}\\
   \widetilde{K}_{22}^{(d)}(\kappa_{a1},\lambda_{b1})&=&\widetilde{Z}_{\frac{1/0}{\overset{}{1/0}}}^{(1)}(\kappa_{a1},\lambda_{b1})-\mathbf{\widetilde{F}}_{d}(\kappa_{a1})\mathbf{\widetilde{M}}_{d}^{-1}\mathbf{\widetilde{F}^{\,(*)}}_{d}(\lambda_{b1})\label{3.29}\,.
\end{eqnarray}
With help of the particular cases $(k_1=k_2=1,l_1=l_2=0)$, $(k_1=k_2=0,l_1=l_2=1)$, $(k_1=l_1=1,k_2=l_2=0)$ and $(k_1=l_1=0,k_2=l_2=1)$, we identify
\begin{eqnarray}
 \widetilde{K}_{21}^{(N)}(\kappa_{a1},\kappa_{b2})&=&\frac{1}{N!\det\mathbf{\widetilde{M}}_{N}}\frac{\widetilde{Z}_{\frac{1/1}{\overset{}{0/0}}}^{(N)}(\kappa_{a1},\kappa_{b2})}{\kappa_{a1}-\kappa_{b2}}\,,\label{3.30}\\
 \widetilde{K}_{12}^{(N)}(\lambda_{b1},\lambda_{a2})&=&\frac{1}{N!\det\mathbf{\widetilde{M}}_{N}}\frac{\widetilde{Z}_{\frac{0/0}{\overset{}{1/1}}}^{(N)}(\lambda_{b1},\lambda_{a2})}{\lambda_{b1}-\lambda_{a2}}\,,\label{3.31}\\
 \widetilde{K}_{22}^{(N-1)}(\kappa_{a1},\lambda_{b1})&=&\frac{1}{N!\det\mathbf{\widetilde{M}}_{N-1}}\widetilde{Z}_{\frac{1/0}{\overset{}{1/0}}}^{(N)}(\kappa_{a1},\lambda_{b1})\,,\label{3.32}\\
 \widetilde{K}_{11}^{(N+1)}(\lambda_{a2},\kappa_{b2})&=&-\frac{1}{N!\det\mathbf{\widetilde{M}}_{N+1}}\widetilde{Z}_{\frac{0/1}{\overset{}{0/1}}}^{(N)}(\lambda_{a2},\kappa_{b2})\,.\label{3.33}
\end{eqnarray}
The normalization constant is fixed by the case $(k_1=k_2=l_1=l_2=0)$
\begin{equation}\label{3.34}
 \widetilde{C}_{N}=\widetilde{Z}_{\frac{0/0}{\overset{}{0/0}}}^{(N)}=N!\det\mathbf{\widetilde{M}}_{N}\,.
\end{equation}
Thus, we find 
\begin{eqnarray}
 \fl&&\widetilde{Z}_{\frac{k_1/k_2}{\overset{}{l_1/l_2}}}^{(N)}(\kappa,\lambda)=\frac{(-1)^{(l_2+k_2)(l_1+k_1-1)/2}N!}{\det^{l_2+k_1-1}\mathbf{\widetilde{M}}_{d}\sqrt{\Ber_{k_1/k_2}^{(2)}(\kappa)}\sqrt{\Ber_{l_1/l_2}^{(2)}(\lambda)}} \times\nonumber\\
 \fl&\times&\det\left[\begin{array}{cc} \displaystyle\left\{-\frac{\widetilde{Z}_{\frac{0/1}{\overset{}{0/1}}}^{(d-1)}(\lambda_{a2},\kappa_{b2})}{(d-1)!}\right\}\underset{1\leq b\leq k_2}{\underset{1\leq a\leq l_2}{\ }} & \displaystyle\left\{\frac{\widetilde{Z}_{\frac{0/0}{\overset{}{1/1}}}^{(d)}(\lambda_{b1},\lambda_{a2})}{d!(\lambda_{b1}-\lambda_{a2})}\right\}\underset{1\leq b\leq l_1}{\underset{1\leq a\leq l_2}{\ }} \\  \displaystyle\left\{\frac{\widetilde{Z}_{\frac{1/1}{\overset{}{0/0}}}^{(d)}(\kappa_{a1},\kappa_{b2})}{d!(\kappa_{a1}-\kappa_{b2})}\right\}\underset{1\leq b\leq k_2}{\underset{1\leq a\leq k_1}{\ }} & \displaystyle\left\{\frac{\widetilde{Z}_{\frac{1/0}{\overset{}{1/0}}}^{(d+1)}(\kappa_{a1},\lambda_{b1})}{(d+1)!}\right\}\underset{1\leq b\leq l_1}{\underset{1\leq a\leq k_1}{\ }}  \end{array}\right]\,.\label{3.35}
\end{eqnarray}
Once more, we notice that the distribution $g(z)$ is quite arbitrary and, thus, a large class of ensembles is covered. The result \eref{3.35} is equivalent to the one found by Bergere \cite{Ber04} with the method of biorthogonal polynomials.

We derive the integral \eref{3.17} for the case that $d=k_2+N-k_1=l_2+N-l_1\geq 0$ is violated by the same method as used in Sec. \ref{sec3.1}. By extending the quotient of characteristic polynomials to the case discussed above, we apply the known result and take the limits with help of l'Hospital's rule. This procedure gives us expressions similar to Eqs.~\eref{3.15} and \eref{3.16}.

For the particular case that $d_\kappa=k_1-k_2-N$ and $d_\lambda=l_1-l_2-N$ are larger than zero, we find a much simpler expression
\begin{eqnarray}
 \fl&&\widetilde{Z}_{\frac{k_1/k_2}{\overset{}{l_1/l_2}}}^{(N)}(\kappa,\lambda)=\frac{(-1)^{(l_1+k_1)(l_1+k_1-1)/2+N(k_2+l_2+1)}N!}{\sqrt{\Ber_{k_1/k_2}^{(2)}(\kappa)}\sqrt{\Ber_{l_1/l_2}^{(2)}(\lambda)}}\times\nonumber\\
 \fl&\times&\det\left[\begin{array}{ccc} 0 & 0 & \left\{\displaystyle\frac{1}{\lambda_{b1}-\lambda_{a2}}\right\}\underset{1\leq b\leq l_1}{\underset{1\leq a\leq l_2}{\ }} \\ 0 & 0 & \left\{\lambda_{b1}^{a-1}\right\}\underset{1\leq b\leq l_1}{\underset{1\leq a\leq d_{\lambda}}{\ }} \\ \left\{\displaystyle\frac{1}{\kappa_{a1}-\kappa_{b2}}\right\}\underset{1\leq b\leq k_2}{\underset{1\leq a\leq k_1}{\ }} &  \left\{\kappa_{a1}^{b-1}\right\}\underset{1\leq b\leq d_{\kappa}}{\underset{1\leq a\leq k_1}{\ }} & \left\{\widetilde{Z}_{\frac{1/0}{\overset{\ }{1/0}}}^{(1)}(\kappa_{a1},\lambda_{b1})\right\}\underset{1\leq b\leq l_1}{\underset{1\leq a\leq k_1}{\ }} \end{array}\right]\,,\label{3.36}
\end{eqnarray}
which is derived in \ref{app4.2}. We see that the whole integral is determined by $\widetilde{Z}_{\frac{1/0}{1/0}}^{(1)}(\kappa_{a1},\lambda_{b1})$ and some algebraic combinations of the ``Cauchy terms'' $1/(\lambda_{b1}-\lambda_{a2})$ and $1/(\kappa_{a1}-\kappa_{b2})$ and of the ``Vandermonde terms'' $\lambda_{b1}^{a-1}$ and $\kappa_{a1}^{b-1}$ if the number of characteristic polynomials in the denominator exceeds a critical value. Here, we emphasize that Eq.~\eref{3.36} has a simpler structure than Eqs.~\eref{3.15}, \eref{3.16} and \eref{3.35} for $d_\kappa\neq d_\lambda$ since there are no limits to perform. Furthermore, the parameters $d_\lambda$ and $d_\kappa$ are independent as long they are larger or equal than zero.

Eqs.~\eref{3.35} and \eref{3.36} are similar to the results found by Uvarov \cite{Uva69,NikUva88}. He studied the transformation behavior of the orthogonal polynomials when the probability distributions differ in a rational function. This shows the connection between his approach and ours.

\section{Applications for integrals of squared--Vandermonde type}\label{sec4}

In Sec. \ref{sec4.1} we apply our method for integrals of squared--Vandermonde type to the example of Hermitian matrices. Since the method has a broad field of applications, we give a list of matrix ensembles in Sec. \ref{sec4.2}.

\subsection{Hermitian matrix ensemble}\label{sec4.1}

We first consider rotation invariant ensembles over the $N\times N$ Hermitian matrices $\Herm(N)$. The averages
\begin{equation}\label{4.1}
 Z_{\tilde{k}_1/\tilde{k}_2}^{(N)}(\kappa)=\int\limits_{\Herm(N)}P(H)\frac{\prod\limits_{j=1}^{\tilde{k}_2}\det(H-\kappa_{j2}\eins_N)}{\prod\limits_{j=1}^{\tilde{k}_1}\det(H-\kappa_{j1}\eins_N)}d[H]
\end{equation}
are of considerable interest. Here, all $\kappa_{j1}$ have an imaginary part.  Equation~\eref{4.1} yields the $k$--point correlation function by differentiation with respect to the source variables $J$ in $\kappa$ \cite{GMW98,Zir06,Guh06,KGG08}, see below. The average over characteristic polynomials in general and their relation to determinantal structures are considered as well, see Refs. \cite{Guh96b,Guh96c,BreHik00,MehNor01,Fyo02,GGK04,BorStr05}.

Let $k_2+l_2=\tilde{k}_2$, $k_1+l_1=\tilde{k}_1$ and $d=k_2+N-k_1=l_2+N-l_1\geq 0$. We consider probability densities $P$ which factorize in the eigenvalue representation of the matrix $H$. Due to the rotation invariance, we diagonalize $H=UEU^\dagger$ with a unitary matrix $U\in\U(N)$. The measure is
\begin{equation}\label{4.2}
 d[H]=\frac{1}{N!}\prod\limits_{j=1}^N\frac{\pi^{j-1}}{(j-1)!}\Delta_N^2(E)d[E]d\mu(U),
\end{equation}
where $d\mu(U)$ is the normalized Haar measure. Thus, we find from Eq.~\eref{3.17}
\begin{equation}\label{4.3}
 \widetilde{Z}_{\frac{k_1/k_2}{\overset{}{l_1/l_2}}}^{(N)}(\tilde{\kappa};\lambda)=(-1)^{(\tilde{k}_1+\tilde{k}_2)N}N!\prod\limits_{j=1}^N\frac{(j-1)!}{\pi^{j-1}}Z_{\tilde{k}_1/\tilde{k}_2}^{(N)}(\kappa)
\end{equation}
with
\begin{equation}\label{4.4}
 g(z_{j})=P(E_j)\delta(y_j)\,.
\end{equation}
We decompose $z_j$ into real and imaginary part, $z_j=E_j+\imath y_j$, and define the two sets $\tilde{\kappa}=\diag(\kappa_{11},\ldots,\kappa_{k_11},\kappa_{12},\ldots,\kappa_{k_22})$ and $\lambda=\diag(\kappa_{k_1+1,1},\ldots,\kappa_{\tilde{k}_11},\kappa_{k_2+1,2},\ldots,\kappa_{\tilde{k}_2,2})$. Our result for this choice, indeed, coincides with the one found by Borodin and Strahov \cite{BorStr05}. They as well splitted the number of characteristic polynomials in two sets and derived the determinantal structure by discrete approximation. They used similar algebraic manipulations but they did not consider the connection to the supersymmetry. Hence our proof is truly a short-cut. As in Ref.~\cite{BorStr05}, the splitting of $\tilde{k}_1$ and of $\tilde{k}_2$ in four positive integers is not unique. Thus, we find different determinantal expressions.

We remark that we have only used the structure of the square roots of the Berezinians and no other property of superspaces. However, we may identify the terms $1/(\kappa_{a1}-\kappa_{b2})$ in Eqs.~\eref{3.27} and \eref{3.28} with the Efetov--Wegner terms \cite{Weg83,Efe83,Guh06} which only appear in superspace. When calculating Eq.~\eref{4.1} with the supersymmetry method, such terms occur by a change of coordinates in superspace from Cartesian coordinates to eigenvalue--angle coordinates \cite{Rot87,KKG08}.

The second term in Eqs.~\eref{3.27} and \eref{3.28}, also contained in Eqs.~\eref{3.26} and \eref{3.29}, are intimately connected to the well known sum over products of orthogonal polynomials. This is borne out in the presence of $\widetilde{\mathbf{M}}_{N}^{-1}$ which generates the bi-orthogonal polynomials \cite{Meh04}. Also, we might choose arbitrary polynomials in the square root of the Berezinian \eref{2.12} instead of the powers $\kappa_{b2}^{a-1}$. If we take the orthogonal polynomials of the probability density $P$, then $\widetilde{\mathbf{M}}_{N}$ becomes diagonal and Eq. \eref{3.26} is indeed the well known result. The $k$--point correlation function can be derived by the case $k_1=k_2=k$ and $l_1=l_2=0$ with $\kappa=\diag(x_1+L_1\imath\varepsilon-J_1,\ldots,x_{k}+L_k\imath\varepsilon-J_k,x_1+L_1\imath\varepsilon+J_1,\ldots,x_{k}+L_k\imath\varepsilon+J_k)$ where $J_j$ are the source variables and $L_j\in\{\pm1\}$ with $\varepsilon$ determine on which side of the real axis the  $\kappa_j$ are. The Cauchy integrals \eref{3.21} and \eref{3.22} become integrals over Dirac distributions by summation over all terms with $L_j=\pm1$ in the limit $\varepsilon\searrow0$. Thus, we find the orthogonal polynomials, too. Due to the differentiation with respect to the $J_j$ at zero the Efetov--Wegner terms vanish and the well known result \cite{Meh04} remains.

\subsection{List of other matrix ensembles}\label{sec4.2}

As we have seen for the ensemble over the Hermitian matrices, we find a determinantal structure \eref{3.35} for particular matrix ensembles with help of the general integral \eref{3.17}. Here, we collect a variety of different matrix ensembles. Those ensembles share not more than two features: (i) the probability density function factorizes in functions of the individual eigenvalues and (ii) the non-factorizing part in the integrand is the squared Vandermonde determinant. We emphasize that this list is not complete. One can certainly find other applications.

We introduce the decomposition into real and imaginary part, $z_j=x_j+\imath y_j$, and the polar coordinates $z_j=r_je^{\imath\varphi_j}$. The probability density $g(z)$ in Eq.~\eref{3.17} for particular ensembles with unitary rotation symmetry is up to constants listed in table \ref{t1}.

\begin{table}[tbp] \centering
\rotatebox{90}{
\begin{tabular}[c]{l|c|c|c}
 matrix ensemble & probability density $P$ &  matrices in the & probability density \\
  & for the matrices & characteristic & $g(z)$ \\
  & & polynomials &  \\
 \noalign{\vskip\doublerulesep\hrule height 2pt}
 Hermitian ensemble  & $\overset{}{\widetilde{P}}\left(\tr H^m,m\in\mathbb{N}\right)$ & $H$ & $P(x)\delta(y)$ \\
 \cite{BreHik00,MehNor01,BreHik03,BDS03,GGK04,BorStr05} & $H=H^\dagger$ & & \\ \hline 
 circular unitary ensemble & $\overset{}{\widetilde{P}}\left(\tr U^m,m\in\mathbb{N}\right)$ & $U$ and $U^\dagger$ & $P\left(e^{\imath\varphi}\right)\delta(r-1)$ \\ 
 (unitary group) & $U^\dagger U=\eins_N$ & & \\
 \cite{BasFor94,Zir96b,KeaSna00,HKC00,CFS05,HPZ05,BumGam06,CFZ07} & & & \\ \hline
 Hermitian chiral (complex & $\overset{\ }{\widetilde{P}}\left(\tr (AA^\dagger)^m,m\in\mathbb{N}\right)$ & $AA^\dagger$ &  $P(x)x^{M-N}\Theta(x)\delta(y)$ \\ 
 Laguerre) ensemble  & $A$ is a complex& &  \\
 \cite{ShuVer93,NagFor95,ADMN97,FyoStr02}& $N\times M$ matrix with $N\leq M$ & & \\ \hline
 Gaussian elliptical ensemble & $\displaystyle\overset{\ }{\exp\left[-\frac{(\tau+1)}{2}\tr H^\dagger H\right]}\times$ & $\overset{\ }{H}$ and $H^{\dagger}$ & $\displaystyle\exp\left[-r^2\left(\sin^2\varphi+\tau\cos^2\varphi\right)\right]$ \\
 \cite{Ake01,Ake02,Ake03,AkeVer03}; for $\tau=1$ & $\displaystyle\times\exp\left[-\frac{(\tau-1)}{2}\RE\,\tr H^2\right]$ &  & \\
 complex Ginibre ensemble & $H$ is a complex matrix; $\tau>0$ & & \\ \hline
 & $\exp\left[-\tr A^\dagger A-\tr B^\dagger B\right]$ & $CD$ and $D^\dagger C^\dagger$ & $\displaystyle\overset{}{K_{M-N}\left(\frac{1+\mu^2}{2\mu^2}r\right)}r^{M-N}\times$\\
 Gaussian complex chiral & $C=\imath A+\mu B$ &  & $\displaystyle\times \exp\left(\frac{1-\mu^2}{2\mu^2}r\cos\varphi\right)$ \\
 ensemble \cite{Osb04} & $D=\imath A^\dagger+\mu B^\dagger$ & & \\
 & $A$ and $B$ are complex $N\times M$ & & \\
  & matrices with $N\leq M$ & &
\end{tabular}}
\caption{\label{t1}  Particular cases of the probability densities $g(z)$ and their corresponding matrix ensembles of unitary rotation symmetry. The joint probability densities are equivalent to the single $g(z)$. $K_{M-N}$ is the modified Bessel function.}
\end{table}

\begin{table}[tbp] \centering
\rotatebox{90}{
\begin{tabular}[c]{l|c|c|c}
 matrix ensemble & probability density $P$ &  matrices in the & probability density  \\
  & for the matrices  & characteristic & $g(z)$ \\
  &   & polynomials &  \\
 \noalign{\vskip\doublerulesep\hrule height 2pt}
 real anti-symmetric matri-  & $\overset{}{\widetilde{P}}\left(\tr H^m,m\in\mathbb{N}\right)$ & $H$ & $P(x)x^{\chi-1/2}\Theta(x)\delta(y)$ \\
 ces (Lie algebra of & $H=-H^T=H^*$ & & \\
 the orthogonal group)\cite{Meh04} & $N=2L+\chi$ dimensional & & \\ \hline 
 special orthogonal & $\overset{}{\widetilde{P}}\left(\tr O^m,m\in\mathbb{N}\right)$ & $O$ & $\displaystyle\overset{}{\frac{P(x)}{\sqrt{1-x^2}}}\delta(y)\left|1-x\right|^{\chi}\times$ \\ 
 group \cite{CFS05,HPZ05,BumGam06} & $O^TO=\eins_{2L+\chi}$ & & $\times\Theta(x-1)\Theta(1-x)$ \\ \hline
 anti-selfdual matrices & $\overset{}{\widetilde{P}}\left(\tr H^m,m\in\mathbb{N}\right)$ & $H$ & $P(x)x^{1/2}\Theta(x)\delta(y)$ \\
 (Lie algebra of the & $H=\left[\begin{array}{cc} 0 & \eins_L \\ -\eins_L & 0 \end{array}\right]H^T\left[\begin{array}{cc} 0 & \eins_L \\ -\eins_L & 0 \end{array}\right]$ & & \\
 unitary--symplectic group) & & & \\ \hline 
 unitary--symplectic group & $\overset{}{\widetilde{P}}\left(\tr S^m,m\in\mathbb{N}\right)$ & $S$ & $\displaystyle\overset{}{P(x)\sqrt{1-x^2}}\delta(y)\times$ \\ 
 \cite{CFS05,HPZ05,BumGam06} & $S^T\left[\begin{array}{cc} 0 & \eins_L \\ -\eins_L & 0 \end{array}\right]S=\left[\begin{array}{cc} 0 & \eins_L \\ -\eins_L & 0 \end{array}\right]$ & & $\times\Theta(x-1)\Theta(1-x)$
\end{tabular}}
\caption{\label{t2}  Particular cases of the probability densities $g(z)$ and their corresponding matrix ensembles of orthogonal and unitary--symplectic rotation symmetry. The joint probability densities are equivalent to the single $g(z)$. The variable $\chi$ is either zero or unity.}
\end{table}

Since the unitary ensembles describe physical systems with broken time reversal symmetry, one is also interested in ensembles which have orthogonal and unitary-symplectic rotation symmetry.  For most of such ensembles the average over ratios of characteristic polynomials can not be transformed to one of the types of integrals discussed in Sec.~\ref{sec3}. However for the special orthogonal group and unitary-symplectic group and the Lie algebras thereof, they are integrals of the squared--Vandermonde type. They are listed in table \ref{t2}.

We remark that the integrals which have to be performed are different for every single matrix ensembles and can be quite difficultly to calculate. Nonetheless all averages over ratios of characteristic polynomials have the determinantal structure \eref{3.35}. The entries of the matrix in the determinant are more or less averages of two characteristic polynomials only. Thus, we achieve a drastic reduction from averages over a large number of characteristic polynomial ratios to averages over two characteristic polynomials for a broad class of random matrix ensembles.

\section{Applications for integrals of square root--Berezinian type}\label{sec5}

In Sec. \ref{sec5.1}, we consider the Hermitian matrices again. We will show that the integral \eref{4.1} can also be understood as an integral of square root--Berezinian type. In Sec. \ref{sec5.2}, we shift this ensemble by an external field $H_0$ and transform the average in the usual way to an integral over a superspace. This integral is an integral of square root--Berezinian type. The determinantal and Pfaffian structure of the $k$--point correlation function for intermediate ensembles is discused in Sec. \ref{sec5.3}.

\subsection{The Hermitian matrices revisited}\label{sec5.1}

The integral \eref{4.1} in eigenvalue--angle coordinates is invariant under permutation of the eigenvalues of the matrix $H$. As in Sec. \ref{sec1.b}, we present one of the Vandermonde determinants as a product over powers of the eigenvalues,
\begin{equation}\label{5.1}
 \fl Z_{\tilde{k}_1/\tilde{k}_2}^{(N)}(\kappa)=\prod\limits_{j=1}^N\frac{(-\pi)^{j-1}}{(j-1)!}\int\limits_{\mathbb{R}^N}\prod\limits_{j=1}^NP(E_j)E_j^{j-1}\frac{\prod\limits_{a=1}^{N}\prod\limits_{b=1}^{\tilde{k}_2}(E_a-\kappa_{b2})}{\prod\limits_{a=1}^{N}\prod\limits_{b=1}^{\tilde{k}_1}(E_a-\kappa_{b1})}\Delta_N(E)d[E]\,.
\end{equation}
Using the same decomposition of $z_j=E_j+\imath y_j$ as in Sec.~\ref{sec4}, we identify this integral with the integral \eref{3.1} and find
\begin{equation}\label{5.2}
 \fl Z_{\tilde{k}_1/\tilde{k}_2}^{(0/N)}(\kappa)=(-1)^{k_1N}\prod\limits_{j=1}^N\frac{(j-1)!}{(-\pi)^{j-1}}Z_{\tilde{k}_1/\tilde{k}_2}^{(N)}(\kappa)
\end{equation}
with
\begin{equation}\label{5.3}
 f_j(z_j)=E_j^{j-1}P(E_j)\delta(y_j)\,.
\end{equation}
The integral \eref{4.1} is permutation invariant with respect to the bosonic and fermionic entries of $\kappa$. However, we do not see this symmetry in the expression found in Sec.~\ref{sec4.1} because we split $\kappa$ into two parts. In the present section, we find a result which shows this symmetry from the beginning, see Eqs. \eref{3.15} and \eref{3.16} for details.

\subsection{The Hermitian matrix ensemble in an external field}\label{sec5.2}

Another calculation of integrals of the squared--Vandermonde type is not the only reason to consider integrals of the square root--Berezinian type. One of its powerful applications is the calculation of the $k$--point correlation functions of ensembles in the presence of an external field. We generalize the result for arbitrary unitarily invariant ensembles over Hermitian matrices in Ref. \cite{Guh06} to ensembles in an external field. Assuming $k\leq N$, we consider the integral
\begin{equation}\label{5.4}
 Z=\int\limits_{\Herm(N)}P(H)\prod\limits_{j=1}^{k}\frac{\det(H+\alpha H_0-\kappa_{j2}\eins_N)}{\det(H+\alpha H_0-\kappa_{j1}\eins_N)}d[H]\,,
\end{equation}
where $P$ is an arbitrary rotation invariant ensemble and $H_0$ is an external field with a coupling constant $\alpha$. For simplicity we set all imaginary parts of $\kappa$ equal to $-\varepsilon$ which means $\kappa=\diag(x_1-\imath\varepsilon-J_1,\ldots,x_{k}-\imath\varepsilon-J_k,x_1-\imath\varepsilon+J_1,\ldots,x_{k}-\imath\varepsilon+J_k)=x^{-}+J$.

We use the generalized Hubbard--Stratonovich transformation \cite{Guh06,KGG08,KSG09} to transform this integral to an integral over supermatrices. With help of the generalized Hubbard--Stratonovich transformation we arrive at
\begin{eqnarray}
 Z&=& 2^{2k(k-1)}\int\limits_{\Sigma_\psi(k)}\int\limits_{\Sigma_{-\psi}(k)}\Phi(\rho)\Sdet^{-1}(\sigma^+\otimes\eins_N+\alpha\eins_{k+k}\otimes H_0)\times\nonumber\\
 &&\times\exp\left[-\imath\Str\rho(\sigma^++\kappa)\right]d[\sigma]d[\rho] \label{5.12}
\end{eqnarray}
with $\sigma^+=\sigma+\imath\varepsilon\eins_{k+k}$. The superfunction $\Phi$ is a rotation invariant supersymmetric extension of the characteristic function
\begin{equation}\label{5.12b}
 \mathcal{F}P(K)=\int\limits_{\Herm(N)}P(H)\exp(\imath\tr HK)d[H]\,.
\end{equation}
Matrices in $\Sigma_0(k)$ are Hermitian and their entries are either of zero or first order in the Grassmann variables. The supermatrix $\rho$ stems from the Wick--rotated set $\Sigma_{\psi}(k)=\Pi_\psi\Sigma_0(k)\Pi_\psi$ with the generalized Wick--rotation $\Pi_\psi=\diag(\eins_{k},e^{\psi/2}\eins_{k})$ with $\psi\in]0,\pi[$. The supermatrix $\sigma$ lies in $\Sigma_{-\psi}(k)$ which is defined accordingly. The Wick--rotation is for norm--dependend ensembles given by $e^{\imath\psi}=\imath$ and guarantees the convergence of the integrals. For cases  such as $\Phi(\rho)=\exp(-\Str\rho^4)$ we need the generalized Wick--rotation. Here, we assume that the probability density $P$ yields a superfunction $\Phi$ with a Wick--rotation regularizing the integral.

We diagonalize the matrices $\rho$ and $\sigma$, $\rho=UrU^{-1}$ and $\sigma=VsV^{-1}$, and integrate over $U$ and $V$ which are in the supergroup $\U(k/k)$. Since we are interested in the $k$--point correlation function $R_k$ of the shifted probability density $P(H-\alpha H_0)$, we omit the Efetov--Wegner terms occurring from this diagonalization because they yield lower order correlation functions. The supergroup integrals are supersymmetric versions of the Itzykson-Zuber integral \cite{Guh91,Guh96,KKG08}. The integral \eref{5.12} reads
\begin{eqnarray}
 \fl Z&=& \frac{1}{(2\pi\imath)^{2k}(k!)^4}\int\limits_{\mathbb{R}^{4k}}\frac{\Phi(r)\Sdet^{-1}(s^+\otimes\eins_N+\alpha\eins_{k+k}\otimes H_0)}{\sqrt{{\rm Ber}_{k/k}^{(2)}(\kappa)}}\det\left[\exp(-\imath r_{a1}s_{b1})\right]_{1\leq a,b\leq k}\times\nonumber\\
 \fl&&\times\det\left[\exp(\imath r_{a2}s_{b2})\right]_{1\leq a,b\leq k}\det\left[\exp(-\imath r_{a1}(x_b-J_b))\right]_{1\leq a,b\leq k}\times\nonumber\\
 \fl&&\times\det\left[\exp\left(\imath e^{\imath\psi}r_{a2}(x_b+J_b)\right)\right]_{1\leq a,b\leq k}\sqrt{{\rm Ber}_{k/k}^{(2)}(s)}d[s]d[r]\,. \label{5.13}
\end{eqnarray}
Using the permutation invariance within the bosonic and fermionic eigenvalues of $r$ and $s$, we find
\begin{eqnarray}
 \fl Z&=& \frac{1}{(2\pi\imath)^{2k}}\int\limits_{\mathbb{R}^{2k}}\frac{\Phi(r)\exp\left[-\imath\Str r\kappa\right]}{\sqrt{{\rm Ber}_{k/k}^{(2)}(\kappa)}}\times\nonumber\\
 \fl&&\times\int\limits_{\mathbb{R}^{2k}}\prod\limits_{a=1}^k\prod\limits_{b=1}^N\frac{e^{-\imath\psi}s_{a2}+\imath\varepsilon+\alpha E_{b}^{(0)}}{s_{a1}+\imath\varepsilon+\alpha E_{b}^{(0)}}\exp\left[-\imath\Str rs^+\right]\sqrt{{\rm Ber}_{k/k}^{(2)}(s)}d[s]d[r]\,. \label{5.14}
\end{eqnarray}
The integration over $s$ is exactly an integral of the square root--Berezinian type \eref{3.1} with the parameters $N_1=N_2=k$, $k_1=0$ and $k_2=N$. In \ref{app5}, we perform the Fourier transform and find
\begin{eqnarray}
 \fl Z&=& \frac{(-1)^{k(k-1)/2}}{\left(2\pi \right)^k}\int\limits_{\mathbb{R}_+^{k}}\int\limits_{\mathbb{R}^k}d[r]\frac{\Phi(r)\exp\left[-\imath\Str r\kappa\right]}{\Delta_N(\alpha E^{(0)})\sqrt{{\rm Ber}_{k/k}^{(2)}(\kappa)}} \times\label{5.15}\\
 \fl&\times&\hspace*{-0.15cm}\det\hspace*{-0.15cm}\left[\begin{array}{cc} \hspace*{-0.25cm}\left\{\displaystyle \frac{r_{a1}^N}{r_{a1}-e^{\imath\psi}r_{b2}}\left(-e^{-\imath\psi}\frac{\partial}{\partial r_{b2}}\right)^{N-1}\right\}\underset{1\leq a,b\leq k}{\ } & \hspace*{-0.25cm}\left\{\displaystyle\imath\sum\limits_{n=N}^{\infty}\frac{1}{n!}\left(\imath\alpha E_b^{(0)}r_{a1}\right)^{n}\right\}\underset{1\leq b\leq N}{\underset{1\leq a\leq k}{\ }} \\ \hspace*{-0.25cm}\left\{\displaystyle2\pi\left(\frac{e^{-\imath\psi}}{\imath}\frac{\partial}{\partial r_{b2}}\right)^{a-1}\right\}\underset{1\leq b\leq k}{\underset{1\leq a\leq N}{\ }} & \hspace*{-0.25cm}\left\{\left(-\alpha E_b^{(0)}\right)^{a-1}\right\}\underset{1\leq a,b\leq N}{\ } \end{array}\right]\hspace*{-0.15cm}\delta(r_2)\nonumber. 
\end{eqnarray}
We notice that the integration domain for the bosonic eigenvalues $r_{a1}$ is the positive real axis where the integral for the fermionic eigenvalues is evaluated at zero.

Indeed, we obtain the known result \cite{Guh06,KGG08} for non--shifted arbitrary unitarily rotation invariant ensembles for $\alpha\to 0$. To show this, we put the $1/\alpha$ terms of the Vandermonde determinant $\Delta_N(\alpha E^{(0)})$ in the last $N$ rows such that the lower right block is independent of $\alpha$. The first $N$ terms of the power series of exponential function in the upper right block are missing. Hence, an expansion in $k$ columns yields that up to one term all other terms are at least of order $\alpha$ at the zero point. We find the limit
\begin{eqnarray}
 \underset{\alpha\to0}{\lim}\,Z&=&  \frac{(-1)^{k(k-1)/2}}{\left(2\pi \right)^k}\int\limits_{\mathbb{R}_+^{k}}\int\limits_{\mathbb{R}^k}\frac{\Phi(r)\exp\left[-\imath\Str r\kappa\right]}{\sqrt{{\rm Ber}_{k/k}^{(2)}(\kappa)}} \times\nonumber\\
 &\times&\det\left[\displaystyle\frac{r_{a1}^N}{r_{a1}-e^{\imath\psi}r_{b2}}\left(-e^{-\imath\psi}\frac{\partial}{\partial r_{b2}}\right)^{N-1}\right]_{1\leq a,b\leq k}\delta(r_2)d[r]\,. \label{5.16}
\end{eqnarray}
By differentiating the source variables in Eq.~\eref{5.15} and setting them to zero, we obtain the modified $k$--point correlation function
\begin{eqnarray}
 \fl &&\widehat{R}_k(x^-)=\left.\prod\limits_{j=1}^k\left(\frac{1}{2}\frac{\partial}{\partial J_j}\right)Z\right|_{J=0}=\nonumber\\
 \fl&=&  \frac{1}{\left(-2\pi \right)^k}\int\limits_{\mathbb{R}_+^{k}}\int\limits_{\mathbb{R}^k}\frac{\Phi(r)\exp\left[-\imath\Str rx^-\right]}{\Delta_N(\alpha E^{(0)})} \times\label{5.17}\\
 \fl&\times&\hspace*{-0.15cm}\det\hspace*{-0.15cm}\left[\begin{array}{cc} \hspace*{-0.25cm}\left\{\displaystyle \frac{r_{a1}^N}{r_{a1}-e^{\imath\psi}r_{b2}}\left(-e^{-\imath\psi}\frac{\partial}{\partial r_{b2}}\right)^{N-1}\right\}\underset{1\leq a,b\leq k}{\ } & \hspace*{-0.25cm}\left\{\displaystyle\imath\sum\limits_{n=N}^{\infty}\frac{1}{n!}\left(\imath\alpha E_b^{(0)}r_{a1}\right)^{n}\right\}\underset{1\leq b\leq N}{\underset{1\leq a\leq k}{\ }} \\ \hspace*{-0.25cm}\left\{\displaystyle2\pi\left(\frac{e^{-\imath\psi}}{\imath}\frac{\partial}{\partial r_{b2}}\right)^{a-1}\right\}\underset{1\leq b\leq k}{\underset{1\leq a\leq N}{\ }} & \hspace*{-0.25cm}\left\{\left(-\alpha E_b^{(0)}\right)^{a-1}\right\}\underset{1\leq a,b\leq N}{\ } \end{array}\right]\hspace*{-0.15cm}\delta(r_2)\nonumber.
\end{eqnarray}
As discussed in Refs.~\cite{Guh06,KGG08}, this correlation function is related to the $k$--point correlation function $R_k$ over the flat Fourier transformation in $x$. Hence, we obtain
\begin{eqnarray}
 \fl &&R_k(x)=\frac{\imath^k}{\left(2\pi \right)^{2k}}\int\limits_{\mathbb{R}^{2k}}\frac{\Phi(r)\exp\left[-\imath\Str rx\right]}{\Delta_N(\alpha E^{(0)})} \times\label{5.18}\\
 \fl&\times&\hspace*{-0.15cm}\det\hspace*{-0.15cm}\left[\begin{array}{cc} \hspace*{-0.25cm}\left\{\displaystyle \frac{r_{a1}^N}{r_{a1}-e^{\imath\psi}r_{b2}}\left(-e^{-\imath\psi}\frac{\partial}{\partial r_{b2}}\right)^{N-1}\right\}\underset{1\leq a,b\leq k}{\ } & \hspace*{-0.25cm}\left\{\displaystyle\sum\limits_{n=N}^{\infty}\frac{\imath}{n!}\left(\imath\alpha E_b^{(0)}r_{a1}\right)^{n}\right\}\underset{1\leq b\leq N}{\underset{1\leq a\leq k}{\ }} \\ \hspace*{-0.25cm}\left\{\displaystyle2\pi\left(\frac{e^{-\imath\psi}}{\imath}\frac{\partial}{\partial r_{b2}}\right)^{a-1}\right\}\underset{1\leq b\leq k}{\underset{1\leq a\leq N}{\ }} & \hspace*{-0.25cm}\left\{\left(-\alpha E_b^{(0)}\right)^{a-1}\right\}\underset{1\leq a,b\leq N}{\ } \end{array}\right]\hspace*{-0.15cm}\delta(r_2)\nonumber.
\end{eqnarray}
We emphasize that this result is exact for any rotation invariant probability density as long as this integral above is existent. It generalizes known results \cite{Guh96b,Guh06b} for norm-dependent ensembles.

\subsection{Determinantal and Pfaffian structures for intermediate ensemble}\label{sec5.3}

In Eq. \eref{5.18}, we easily see that for factorizing characteristic function \eref{5.12b} and, thus, for factorizing superfunction
\begin{equation}\label{5.19}
 \Phi(r)=\prod\limits_{j=1}^k\frac{\Phi(r_{j1})}{\Phi(r_{j2})}
\end{equation}
the $k$--point correlation function is a ratio of a $(k+N)\times(k+N)$ determinant and a $N\times N$ determinant
\begin{eqnarray}
 \fl R_k(x)= \frac{\imath^k}{\left(2\pi \right)^{2k}\Delta_N(\alpha E^{(0)})}\det\left[\begin{array}{cc} \left\{\widetilde{R}_1(x_a,x_b)\right\}\underset{1\leq a,b\leq k}{\ } & \left\{\widetilde{R}_2(\alpha E_b^{(0)},x_a)\right\}\underset{1\leq b\leq N}{\underset{1\leq a\leq k}{\ }} \\ \left\{\widetilde{R}_{a3}(x_b)\right\}\underset{1\leq b\leq k}{\underset{1\leq a\leq N}{\ }} & \left\{\left(-\alpha E_b^{(0)}\right)^{a-1}\right\}\underset{1\leq a,b\leq N}{\ } \end{array}\right]\,. \label{5.20}
\end{eqnarray}
Here, the entries are
\begin{eqnarray}
 \fl\widetilde{R}_1(x_a,x_b)&=&\displaystyle\int\limits_{\mathbb{R}^2}\exp\left[-\imath (r_1x_{a}-e^{\imath\psi}r_2x_b)\right]\frac{\Phi(r)r_{1}^N}{r_{1}-e^{\imath\psi}r_{2}}\left(-e^{-\imath\psi}\frac{\partial}{\partial r_{2}}\right)^{N-1}\delta(r_2)d[r]\,,\label{5.21}\\
 \fl\widetilde{R}_2(\alpha E_b^{(0)},x_a)&=&\displaystyle\imath\int\limits_{\mathbb{R}}\Phi(r_1)\exp\left[-\imath r_1x_a\right]\sum\limits_{n=N}^{\infty}\frac{1}{n!}\left(\imath\alpha E_b^{(0)}r_{1}\right)^{n}dr_1\,,\label{5.22}\\
 \fl\widetilde{R}_{a3}(x_b)&=&\displaystyle2\pi\int\limits_{\mathbb{R}}\frac{1}{\Phi\left(e^{\imath\psi}r_2\right)}\exp\left[\imath e^{\imath\psi} r_2x_b\right]\left(\frac{e^{-\imath\psi}}{\imath}\frac{\partial}{\partial r_{2}}\right)^{a-1}\delta(r_2)dr_2\,.\label{5.23}
\end{eqnarray}
By splitting off the lower right block from the determinant as in Eq.~\eref{1.7b}, we see that $R_k$ is a $k\times k$ determinant in $x$ which was also shown in Ref.~\cite{Guh96b}. However, the representation \eref{5.20} is much better suited for further calculations than for the $k\times k$ determinant representation.

We can transform the characteristic function $\Phi$ in Eqs.~\eref{5.21}, \eref{5.22} and \eref{5.23} to probability densities in an ordinary space by the inverse procedure performed in Sec.~\ref{sec5.2}. We emphasize that these correlation functions can also be expressed in terms of mean values over ratios of characteristic polynomials. To illustrate this we explicitly work out Eqs.~\eref{5.21}, \eref{5.22} and \eref{5.23} for Laguerre ensembles in \ref{app6}. For Gaussian ensembles we obtain the correct result \cite{Guh06b}.

Instead of taking $H_0$ as a constant external field, one can take it from a random matrix ensemble, too. For the Gaussian case this was discussed in Refs. \cite{PanMeh83,MehPan83,Guh96b,Guh96c}. Here, we investigate $H_0$ as a real symmetric, a Hermitian and a Hermitian selfdual matrix with factorizing probability density $\widetilde{P}$. We remark that $H_0$ can also be drawn from a Wishart ensemble since it can be mapped to one of the symmetric ensembles.

Assuming that the characteristic function of $P$ factorizes as well, the $k$--point correlation function is
\begin{equation}\label{5.24}
 R_{k}^{(2)}(x)=\frac{(-1)^{N(N-1)/2}\imath^k}{\left(2\pi \right)^{2k}}\det\left[K^{(2)}(x_a,x_b)\right]_{1\leq a,b\leq k}
\end{equation}
for a second ensemble over the Hermitian matrices. We define the kernel
\begin{eqnarray}
 \fl K^{(2)}(x_a,x_b)&=&\widetilde{R}_1(x_a,x_b)-\label{5.25}\\
 \fl &-&\sum\limits_{m,n=1}^N\int\limits_{\mathbb{R}}\widetilde{P}(E)\widetilde{R}_2(\alpha E,x_a)\left(- E\right)^{m-1}dE\left(M^{(2)\,-1}\right)_{mn}\frac{\widetilde{R}_{n3}(x_b)}{\alpha^{n-1}}\nonumber
\end{eqnarray}
and the moment matrix
\begin{equation}\label{5.26}
 M^{(2)}_{mn}=\int\limits_{\mathbb{R}}\widetilde{P}(E)(-E)^{m+n-2}dE
\end{equation}
for probability density $\widetilde{P}$.

For quaternionic $H_0$, $N=2Q$, we apply a generalization of de Bruijn's integral theorem \cite{Bru55} which we derive in \ref{app3.2}. This yields the Pfaffian structure
\begin{equation}\label{5.27}
  R_{k}^{(4)}(x)=\frac{\imath^k}{\left(2\pi \right)^{2k}}\Pf\left[\begin{array}{cc} K_{1}^{(4)}(x_a,x_b) & K_{2}^{(4)}(x_a,x_b) \\ -K_{2}^{(4)}(x_b,x_a) & K_{3}^{(4)}(x_a,x_b) \end{array}\right]_{1\leq a,b\leq k}\,,
\end{equation}
where the sign of the Pfaffian for an arbitrary antisymmetric $2N\times2N$ matrix $\{D_{ab}\}$ is defined as
\begin{eqnarray}
 \Pf\left[D_{ab}\right]_{1\leq a,b\leq 2N}=\frac{1}{2^NN!}\sum\limits_{\omega\in\mathfrak{S}_{2N}}\sign(\omega)\prod\limits_{j=1}^{N}D_{\omega(2j)\omega(2j+1)}\,.\label{5.31}
\end{eqnarray}
The set $\mathfrak{S}_M$ is the permutation group over $M$ elements and the sign function yields ``$+1$'' and ``$-1$'' for even and odd permutations, respectively. The kernels in the Pfaffian are given by
\begin{eqnarray}
  &&K_{1}^{(4)}(x_a,x_b)=\sum\limits_{m,n=1}^{2Q}\frac{\widetilde{R}_{m3}(x_a)}{\alpha^{m-1}}\left(M^{(4)\,-1}\right)_{mn}\frac{\widetilde{R}_{n3}(x_b)}{\alpha^{n-1}}\label{5.28}\,,\\
  &&K_{2}^{(4)}(x_a,x_b)=\widetilde{R}_1(x_b,x_a)+\sum\limits_{m,n=1}^{2Q}\frac{\widetilde{R}_{m3}(x_a)}{\alpha^{m-1}}\left(M^{(4)\,-1}\right)_{mn}k_n^{(4)}(x_b)\,,\label{5.29}\\
  &&K_{3}^{(4)}(x_a,x_b)=k^{(4)}(x_a,x_b)+\sum\limits_{m,n=1}^{2Q}k_m^{(4)}(x_a)\left(M^{(4)\,-1}\right)_{mn}k_n^{(4)}(x_b)\,.\label{5.30}
\end{eqnarray}
The functions appearing in these definitions are
\begin{eqnarray}
 \fl k^{(4)}(x_a,x_b)&=&\int\limits_{\mathbb{R}}\widetilde{P}(E)\det\left[\begin{array}{cc}\widetilde{R}_2(\alpha E,x_b) & \widetilde{R}_2(\alpha E,x_a) \\ \displaystyle\frac{\partial \widetilde{R}_2}{\partial E}(\alpha E,x_b) & \displaystyle\frac{\partial \widetilde{R}_2}{\partial E}(\alpha E,x_a) \end{array}\right]dE\,,\label{h5.1}\\
 \fl k_n^{(4)}(x_b)&=&(-1)^n\int\limits_{\mathbb{R}}\widetilde{P}(E)\det\left[\begin{array}{cc} \widetilde{R}_2(\alpha E,x_b) & E^{n-1} \\ \displaystyle\frac{\partial \widetilde{R}_2}{\partial E}(\alpha E,x_b) & (n-1)E^{n-2} \end{array}\right]dE\,.\label{h5.2}
\end{eqnarray}
In Eqs. \eref{5.28}, \eref{5.29} and \eref{5.30} the inverse of the skew--symmetric moment matrix
\begin{equation}\label{5.34}
 M^{(4)}_{ab}=(b-a)\int\limits_{\mathbb{R}}\widetilde{P}(E)(-E)^{a+b-3}dE
\end{equation}
arises which generates the skew orthogonal polynomials of quaternion type.

If $H_0$ stems from an ensemble of $(2Q+\chi)\times(2Q+\chi)$ real symmetric matrices, $\chi\in\{0,1\}$, we obtain another Pfaffian
\begin{equation}
  R_{k}^{(1)}(x)=\frac{(-1)^{N(N-1)/2}\imath^k}{\left(2\pi \right)^{2k}}\Pf\left[\begin{array}{cc} K_{1}^{(1)}(x_a,x_b) & K_{2}^{(1)}(x_a,x_b) \\ -K_{2}^{(1)}(x_b,x_a) & K_{3}^{(1)}(x_a,x_b) \end{array}\right]_{1\leq a,b\leq k}\hspace*{-0.5cm}.\label{5.35}
\end{equation}
The entries are
\begin{eqnarray}
  &&K_{1}^{(1)}(x_a,x_b)=\sum\limits_{m,n=1}^{2Q}\frac{\widetilde{R}_{m3}(x_a)}{\alpha^{m-1}}\left(M^{(1)\,-1}\right)_{mn}\frac{\widetilde{R}_{n3}(x_b)}{\alpha^{n-1}}\label{5.36}\,,\\
  &&K_{2}^{(1)}(x_a,x_b)=\widetilde{R}_1(x_b,x_a)+\sum\limits_{m,n=1}^{2Q}\frac{\widetilde{R}_{m3}(x_a)}{\alpha^{m-1}}\left(M^{(1)\,-1}\right)_{mn}k_n^{(1)}(x_b)\,,\label{5.37}\\
  &&K_{3}^{(1)}(x_a,x_b)=k^{(1)}(x_a,x_b)+\sum\limits_{m,n=1}^{2Q}k_m^{(1)}(x_a)\left(M^{(1)\,-1}\right)_{mn}k_n^{(1)}(x_b)\label{5.38}
\end{eqnarray}
with the moment matrix
\begin{equation}\label{5.39}
 \fl M^{(1)}_{mn}=\left\{\begin{array}{ll} \hspace*{-0.2cm}\int\limits_{-\infty<E_1<E_2<\infty}\hspace*{-0.7cm}\widetilde{P}(E)\det\left[\begin{array}{cc}(-E_1)^{b-1} & (-E_1)^{a-1} \\ (-E_2)^{b-1} & (-E_2)^{a-1} \end{array}\right]d[E] & \hspace*{-0.2cm},\,1\leq m,n\leq2Q+\chi \\
 & \\
 \hspace*{-0.2cm}-\int\limits_{\mathbb{R}}\widetilde{P}(E)(-E)^{m-1}dE & \hspace*{-0.2cm},\left\{\begin{array}{l}\,1\leq m\leq2Q \\
   \,n=2Q+2 \\ \chi=1 \end{array}\right.\\
 \hspace*{-0.2cm}\int\limits_{\mathbb{R}}\widetilde{P}(E)(-E)^{n-1}dE & \hspace*{-0.2cm},\left\{\begin{array}{l}\,1\leq n\leq2Q \\
   \,m=2Q+2 \\ \chi=1 \end{array}\right.\\
 \hspace*{-0.2cm}0 & \hspace*{-0.2cm},\left\{\begin{array}{l}\,m=n=2Q+2 \\ \chi=1\end{array}\right.\end{array}\right.\hspace*{-0.5cm}.
\end{equation}
Here, the functions in Eqs. \eref{5.37} and \eref{5.38} are
\begin{equation}
 \fl k^{(1)}(x_a,x_b)=\int\limits_{-\infty<E_1<E_2<\infty}\widetilde{P}(E)\det\left[\begin{array}{cc}\widetilde{R}_2(\alpha E_2,x_a) & \widetilde{R}_2(\alpha E_2,x_b) \\ \widetilde{R}_2(\alpha E_1,x_a) & \widetilde{R}_2(\alpha E_1,x_b) \end{array}\right]d[E]\label{5.40}
\end{equation}
and
\begin{eqnarray}
 \fl &&\left[k_n^{(1)}(x_b)\right]_{1\leq n\leq2(Q+\chi)}=\nonumber\\
 \fl&=&\left[\begin{array}{c} \left\{\int\limits_{-\infty<E_1<E_2<\infty}\hspace*{-1cm}\widetilde{P}(E)\det\left[\begin{array}{cc} \widetilde{R}_2(\alpha E_2,x_b) & (-E_2)^{n-1} \\ \widetilde{R}_2(\alpha E_1,x_b) & (-E_1)^{n-1} \end{array}\right]d[E]\right\}_{1\leq n\leq2Q+\chi} \\ 
 -\int\limits_{\mathbb{R}}\widetilde{P}(E)\widetilde{R}_2(\alpha E,x_b)dE \end{array}\right].\label{5.41}
\end{eqnarray}
As in the other cases, the matrix $\{M_{mn}\}$ generates the skew orthogonal polynomials of real type with respect to $\widetilde{P}$. Pandey and Mehta \cite{PanMeh83} constructed these polynomials for the Gaussian measure. They also found a Pfaffian structure for the interpolation between GUE and GOE. In Ref.~\cite{Guh96b}, one can implicitly recognize the determinantal and Pfaffian structure in the interpolation from an arbitrary Gaussian symmetric ensemble to GUE. Our results \eref{5.24}, \eref{5.27} and \eref{5.35} extend the determinantal and Pfaffian structures in the intermediate ensembles from an arbitrary symmetric ensemble factorizing in the probability density to an arbitrary unitarily invariant ensemble factorizing in the characteristic function.

Moreover, we can omit the factorization of the unitarily rotation invariant ensemble and find an integral representation in the superspace for an arbitrary unitarily invariant ensemble for an interpolation to the other classes of rotation invariance. For this purpose we integrate over $H_0$ in Eq.~\eref{5.18} and find for the integral kernel a determinant or a Pfaffian determinant, depending on whether $H_0$ is Hermitian, Hermitian self-dual or real symmetric.

\section{Remarks and conclusions}\label{sec6}

We presented a new method to calculate mean values for ratios of characteristic polynomial in a wide class of matrix ensembles with unitary symmetry and factorizing probability density. This method is also applicable to the classical Lie groups and their algebras. Our approach is based on determinantal structures of Berezinians with arbitrary dimensions resulting from diagonalization of symmetric supermarices. Although we did not use any supersymmetry for ordinary matrix ensembles, we managed to reconstruct those Berezinians in the product of the characteristic polynomials with powers of the Vandermonde determinant. Using these determinantal structures, we obtained determinants whose entries are given in terms of the inverse of the moment matrix for the particular ensemble. These matrices are connected to the orthogonal polynomials and show that the known results from the method of orthogonal polynomials \cite{BreHik00c,MehNor01,AkeVer03,Ber04,Meh04} are obtained. In particular, we re-derived the results of Borodin and Strahov \cite{BorStr05} for the ensembles over the symmetric spaces in a more direct way.

The determinantal structure is stable when an external field is coupled to the random matrix. With help of the supersymmetry method, we have shown this for arbitrary unitarily invariant Hermitian matrix ensembles in an external field. Our formula for the $k$--point correlation function is a generalization of recent results over arbitrary Hermitian matrix ensembles \cite{Guh06} and over norm--dependent ensembles in an external field \cite{Guh06b}. Moreover, we considered an external field drawn from another symmetric ensemble. We calculated the $k$--point correlation function for an interpolation between an arbitrary Hermitian ensemble factorizing in the characteristic function and an arbitrary symmetric ensemble factorizing in the probability density. We found determinantal and Pfaffian structures, too. For Gaussian ensembles, this coincides with known results \cite{PanMeh83,Guh96b}. We gave explicit results for the Laguerre ensembles coupled to an external field in a way which is different from couplings investigated in Ref.~\cite{GuhWet97,DesFor06}.

\section*{Acknowledgements}
We thank G. Akemann, H. Kohler, V. Osipov, M.J. Phillips, H.-J. Sommers and M.R. Zirnbauer for fruitful discussions and for important references. We acknowledge support from Deutsche Forschungsgemeinschaft within Sonderforschungsbereich Transregio 12 ``Symmetries and Universality in Mesoscopic Systems''.

\appendix

\section{Derivation of the Berezinians}\label{app1}

In \ref{app1.1} we derive the determinantal structure for $\beta=2$. The derivations for the other two cases $p\leq2q$ and $p\geq2q$ for $\beta\in\{1,4\}$ are given in \ref{app1.2} and in \ref{app1.3}, respectively.

\subsection{The case $\beta=2$}\label{app1.1}

Let $p\leq q$. The trick is to extend the ratio of products by additional variables $\diag(\kappa_{p+1,1},\ldots,\kappa_{q,1})$. Then, we apply Eq. \eref{2.3}. Since these additional variables are artificially introduced and, hence, arbitrary, we perform the limit $\diag(\kappa_{p+1,1},\ldots,\kappa_{q,1})\to 0$. We find
\begin{eqnarray}
 \fl & &\frac{\prod\limits_{1\leq a<b\leq p}\left(\kappa_{a1}-\kappa_{b1}\right)\prod\limits_{1\leq a<b\leq q}\left(\kappa_{a2}-\kappa_{b2}\right)}{\prod\limits_{a=1}^p\prod\limits_{b=1}^q\left(\kappa_{a1}-\kappa_{b2}\right)}=\nonumber\\
 \fl & = &\frac{\prod\limits_{a=p+1}^q\prod\limits_{b=1}^q\left(\kappa_{a1}-\kappa_{b2}\right)}{\prod\limits_{p+1\leq a<b\leq q}\left(\kappa_{a1}-\kappa_{b1}\right)\prod\limits_{a=1}^p\prod\limits_{b=p+1}^q\left(\kappa_{a1}-\kappa_{b1}\right)}\frac{\prod\limits_{1\leq a<b\leq q}\left(\kappa_{a1}-\kappa_{b1}\right)\left(\kappa_{a2}-\kappa_{b2}\right)}{\prod\limits_{1\leq a,b\leq q}\left(\kappa_{a1}-\kappa_{b2}\right)}=\nonumber\\
 \fl & = & \left.\frac{(-1)^{q(q+1)/2+pq}\prod\limits_{a=1}^q\prod\limits_{b=p+1}^q\left(\kappa_{a2}-\kappa_{b1}\right)}{\prod\limits_{p+1\leq a<b\leq q}\left(\kappa_{a1}-\kappa_{b1}\right)\prod\limits_{a=1}^p\prod\limits_{b=p+1}^q\left(\kappa_{a1}-\kappa_{b1}\right)}\det\left[\frac{1}{\kappa_{a1}-\kappa_{b2}}\right]_{1\leq a,b\leq q}\right|_{\kappa_{p+1,1}=\ldots=\kappa_{q,1}=0}\hspace*{-1.5cm}=\nonumber\\
 \fl & = & \left.(-1)^{q(q+1)/2+pq}\frac{\displaystyle\det\left[\frac{1}{\kappa_{a1}-\kappa_{b2}}\right]_{1\leq a,b\leq q}}{\prod\limits_{p+1\leq a<b\leq q}\left(\kappa_{a1}-\kappa_{b1}\right)}\right|_{\kappa_{p+1,1}=\ldots=\kappa_{q,1}=0}\Sdet^{p-q}\kappa\label{a1.1}\,.
\end{eqnarray}
This yields the result \eref{2.7}.

\subsection{The case $\beta\in\{1,4\}$ with $p\leq2q$}\label{app1.2}

Let $p\leq 2q$. This calculation is similar to the one in \ref{app1.1}. The only difference is that we have to take Kramers' degeneracy in the fermionic eigenvalues $\diag(\kappa_{12},\ldots,\kappa_{q2})$ into account. We first use non-degenerate entries to reduce the problem to the one for $\beta=2$. Finally, we restore Kramers' degeneracy. We find
\begin{eqnarray}
 \fl & & \frac{\prod\limits_{1\leq a<b\leq p}\left(\kappa_{a1}-\kappa_{b1}\right)\prod\limits_{1\leq a<b\leq q}\left(\kappa_{a2}-\kappa_{b2}\right)^4}{\prod\limits_{a=1}^p\prod\limits_{b=1}^q\left(\kappa_{a1}-\kappa_{b2}\right)^2}=\nonumber\\
 \fl & = &(-1)^{q(q-1)/2}\left.\frac{\prod\limits_{1\leq a<b\leq p}\left(\kappa_{a1}-\kappa_{b1}\right)\prod\limits_{1\leq a<b\leq 2q}\left(\kappa_{a2}-\kappa_{b2}\right)}{\prod\limits_{a=1}^p\prod\limits_{b=1}^{2q}\left(\kappa_{a1}-\kappa_{b2}\right)\prod\limits_{j=1}^q\left(\kappa_{j2}-\kappa_{j+q,2}\right)}\right|_{\kappa_{j2}=\kappa_{j+q,2}}=\nonumber\\
 \fl & = & \left.(-1)^{q(q+1)/2+p}\prod\limits_{j=1}^q\frac{1}{\left(\kappa_{j2}-\kappa_{j+q,2}\right)}\det\left[\begin{array}{c}\left\{\displaystyle\frac{\kappa_{a1}^{p-2q}\kappa_{b2}^{2q-p}}{\kappa_{a1}-\kappa_{b2}}\right\}\underset{1\leq b\leq 2q}{\underset{1\leq a\leq p}{\ }} \\ \left\{\kappa_{b2}^{a-1}\right\}\underset{1\leq b\leq 2q}{\underset{1\leq a\leq 2q-p}{\ }} \end{array}\right]\right|_{\kappa_{j2}=\kappa_{j+q,2}}\label{a1.2}\,.
\end{eqnarray}
This directly leads to Eq.~\eref{2.8}.

\subsection{The case $\beta\in\{1,4\}$ with $p\geq2q$}\label{app1.3}

Let $p\geq 2q$. Some modifications of the line of arguing in \ref{app1.2} are necessary, we find
\begin{eqnarray}
 \fl & & \frac{\prod\limits_{1\leq a<b\leq p}\left(\kappa_{a1}-\kappa_{b1}\right)\prod\limits_{1\leq a<b\leq q}\left(\kappa_{a2}-\kappa_{b2}\right)^4}{\prod\limits_{a=1}^p\prod\limits_{b=1}^q\left(\kappa_{a1}-\kappa_{b2}\right)^2}=\nonumber\\
 \fl & = &(-1)^{q(q-1)/2}\left.\frac{\prod\limits_{1\leq a<b\leq p}\left(\kappa_{a1}-\kappa_{b1}\right)\prod\limits_{1\leq a<b\leq 2q}\left(\kappa_{a2}-\kappa_{b2}\right)}{\prod\limits_{a=1}^p\prod\limits_{b=1}^{2q}\left(\kappa_{a1}-\kappa_{b2}\right)\prod\limits_{j=1}^q\left(\kappa_{j2}-\kappa_{j+q,2}\right)}\right|_{\kappa_{j2}=\kappa_{j+q,2}}=\nonumber\\
 \fl & = & (-1)^{[p(p-1)+q(q-1)]/2}\times\nonumber\\
 \fl &\times& \left.\prod\limits_{j=1}^q\frac{(\kappa_{j2}\kappa_{j+q,2})^{2q-p}}{\left(\kappa_{j2}-\kappa_{j+q,2}\right)}\det\left[\begin{array}{cc}\left\{\displaystyle\frac{\kappa_{a1}^{p-2q}}{\kappa_{a1}-\kappa_{b2}}\right\}\underset{1\leq b\leq 2q}{\underset{1\leq a\leq p}{\ }} & \left\{\kappa_{a2}^{b-1}\right\}\underset{1\leq b\leq p-2q}{\underset{1\leq a\leq p}{\ }} \end{array}\right]\right|_{\kappa_{j2}=\kappa_{j+q,2}}\label{a1.3}\,,
\end{eqnarray}
which implies formula \eref{2.9}.

\section{Calculations of integrals of square root--Berezinian type}\label{app2}

\subsection{The case $k_1=k_2=k$}\label{app2.1}

We calculate
\begin{eqnarray}
 \fl&&\int\limits_{\mathbb{C}^{N_1+N_2}}\prod\limits_{j=1}^{N_1}g_j(z_{j1})\prod\limits_{j=1}^{N_2}f_j(z_{j2})\sqrt{\Ber_{N_1+k/N_2+k}^{(2)}(\tilde{z})}d[z]=(-1)^{(N_2+k)(N_2+k-1)/2}\times\label{a2.1}\\
 \fl&\times&
  \int\limits_{\mathbb{C}^{N_1+N_2}}\prod\limits_{j=1}^{N_1}g_j(z_{j1})\prod\limits_{j=1}^{N_2}f_j(z_{j2})\det\left[\begin{array}{cc}  \left\{\displaystyle\frac{1}{\kappa_{a1}-\kappa_{b2}}\right\}\underset{1\leq a,b\leq k}{\ } & \left\{\displaystyle\frac{1}{\kappa_{a1}-z_{b2}}\right\}\underset{1\leq b\leq N_2}{\underset{1\leq a\leq k}{\ }} \\  \left\{\kappa_{b2}^{a-1}\right\}\underset{1\leq b\leq k}{\underset{1\leq a\leq N_2-N_1}{\ }} & \left\{z_{b2}^{a-1}\right\}\underset{1\leq b\leq N_2}{\underset{1\leq a\leq N_2-N_1}{\ }} \\  \left\{\displaystyle\frac{1}{z_{a1}-\kappa_{b2}}\right\}\underset{1\leq b\leq k}{\underset{1\leq a\leq N_1}{\ }} & \left\{\displaystyle\frac{1}{z_{a1}-z_{b2}}\right\}\underset{1\leq b\leq N_2}{\underset{1\leq a\leq N_1}{\ }} \end{array}\right]d[z]\,.\nonumber
\end{eqnarray}
We use the definitions \eref{3.5} to \eref{3.10}. The integral \eref{a2.1}, then, reads
\begin{eqnarray}
 \fl&&\int\limits_{\mathbb{C}^{N_1+N_2}}\prod\limits_{j=1}^{N_1}g_j(z_{j1})\prod\limits_{j=1}^{N_2}f_j(z_{j2})\sqrt{\Ber_{N_1+k/N_2+k}^{(2)}(\tilde{z})}d[z]=\nonumber\\
  \fl&=& (-1)^{(N_2+k)(N_2+k-1)/2}
  \det\left[\begin{array}{cc}  \left\{\displaystyle\frac{1}{\kappa_{a1}-\kappa_{b2}}\right\}\underset{1\leq a,b\leq k}{\ } & \left\{\mathbf{F}^{(N_2)}(\kappa_{a1})\right\}\underset{1\leq a\leq k}{\ } \\  \left\{\mathbf{G}^{(N_1/N_2)}(\kappa_{b2})\right\}\underset{1\leq b\leq k}{\ } & \mathbf{M}_{N_1/N_2} \end{array}\right]\,.\label{a2.2}
\end{eqnarray}
The next step is to extract the matrix $\mathbf{M}_{N_1/N_2}$ from the determinant. This yields
\begin{eqnarray}
 \fl&&\int\limits_{\mathbb{C}^{N_1+N_2}}\prod\limits_{j=1}^{N_1}g_j(z_{j1})\prod\limits_{j=1}^{N_2}f_j(z_{j2})\sqrt{\Ber_{N_1+k/N_2+k}^{(2)}(\tilde{z})}d[z]=\nonumber\\
 \fl&=& (-1)^{(N_2+k)(N_2+k-1)/2} \det\mathbf{M}_{N_1/N_2}
  \det\left[K^{(N_1/N_2)}(\kappa_{a1},\kappa_{b2})\right]\underset{1\leq a,b\leq k}{\ }\label{a2.3}
\end{eqnarray}
with $K^{(N_1/N_2)}$ as in definition \eref{3.11}.

\subsection{The case $k_1\leq k_2=k$}\label{app2.2}

Let $k_1\leq k_2$. Then, we have
\begin{equation}\label{a2.4}
 Z_{k_1/k_2}^{(N_1/N_2)}(\kappa)=(-1)^{(k_2-k_1)N_1}\underset{\kappa_{k_1+1,1},\ldots,\kappa_{k_2,1}\to\infty}{\lim}\prod\limits_{j=k_1+1}^{k_2}\kappa_{j1}^{N_2-N_1}Z_{k_2/k_2}^{(N_1/N_2)}(\kappa)
\end{equation}
With help of Eq. \eref{3.14}, we obtain
\begin{eqnarray}
 \fl&&\displaystyle Z_{k_1/k_2}^{(N_1/N_2)}(\kappa)=\frac{(-1)^{k_1(k_1-1)/2+(k_2-k_1)N_1}}{C_{N_1/N_2}^{k_2-1}\sqrt{\Ber_{k_1/k_2}^{(2)}(\kappa)}}\times\label{a2.5}\\
 \fl&\times&\underset{\kappa_{k_1+1,1},\ldots,\kappa_{k_2,1}\to\infty}{\lim}\prod\limits_{j=k_1+1}^{k_2}\kappa_{j1}^{N_2-N_1+k_2-k_1}\frac{\displaystyle\det\left[\frac{Z_{1/1}^{(N_1/N_2)}(\kappa_{a1},\kappa_{b2})}{\kappa_{a1}-\kappa_{b2}}\right]_{1\leq a,b\leq k_2}}{\det\left[\kappa_{b1}^{a-1}\right]\underset{k_1+1\leq b\leq k_2}{\underset{1\leq a\leq k_2-k_1}{\ }}}\nonumber
\end{eqnarray}
The limit expression is a function of $1/\kappa_{a1}$ which is differentiable at $1/\kappa_{a1}=0$. Hence, using l'Hospital's rule we find Eq.~\eref{3.15}

\subsection{The case $k=k_1\geq k_2$}\label{app2.3}

For $k_1\geq k_2$, we have to proceed in a similar way. We extend the number of the fermionic eigenvalues $\kappa_2$ and find
\begin{eqnarray}
 \fl&& Z_{k_1/k_2}^{(N_1/N_2)}(\kappa)=(-1)^{(k_1-k_2)(N_2-N_1)}\underset{\kappa_{k_2+1,2},\ldots,\kappa_{k_1,2}\to\infty}{\lim}\prod\limits_{j=k_2+1}^{k_1}\kappa_{j2}^{N_1-N_2}Z_{k_1/k_1}^{(N_1/N_2)}(\kappa)=\nonumber\\
 \fl&=&\frac{(-1)^{(k_2+2k_1)(k_2-1)/2+(k_1-k_2)(N_2-N_1)}}{C_{N_1/N_2}^{k_1-1}\sqrt{\Ber_{k_1/k_2}^{(2)}(\kappa)}}\times\nonumber\\
 \fl&\times&\underset{\kappa_{k_1+1,1},\ldots,\kappa_{k_2,1}\to\infty}{\lim}\prod\limits_{j=k_2+1}^{k_1}\kappa_{j2}^{N_1-N_2+k_1-k_2}\frac{\displaystyle\det\left[\frac{Z_{1/1}^{(N_1/N_2)}(\kappa_{b1},\kappa_{a2})}{\kappa_{b1}-\kappa_{a2}}\right]_{1\leq a,b\leq k_1}}{\det\left[\kappa_{b2}^{a-1}\right]\underset{k_2+1\leq b\leq k_1}{\underset{1\leq a\leq k_1-k_2}{\ }}}\,.\label{a2.7}
\end{eqnarray}
This directly gives the result \eref{3.16}.

\section{Extension of integration theorems for determinantal kernels}\label{app3}

\subsection{Extension of Andr\'{e}ief's integral theorem}\label{app3.1}

We consider the integral
\begin{equation}\label{a3.1}
 \fl \mathcal{I}=\int\limits_{\mathbb{C}^N}\det\left[\begin{array}{c} \{r_{ab}\}\underset{1\leq b\leq N+k}{\underset{1\leq a\leq k}{\ }} \\ \{R_b(z_a,z_a^*)\}\underset{1\leq b\leq N+k}{\underset{1\leq a\leq N}{\ }} \end{array}\right]\det\left[\begin{array}{c} \{s_{ab}\}\underset{1\leq b\leq N+l}{\underset{1\leq a\leq l}{\ }} \\ \{S_b(z_a,z_a^*)\}\underset{1\leq b\leq N+l}{\underset{1\leq a\leq N}{\ }} \end{array}\right]d[z]\,.
\end{equation}
The functions $R_a$ and $S_a$ are such that the integrals are convergent. Apart from this property they are arbitrary. We expand the first determinant in the first $k$ rows and the second determinant in the first $l$ rows and obtain
\begin{eqnarray}
 \fl \mathcal{I}&=&\frac{1}{k!(N-k)!l!(N-l)!}\underset{\sigma\in\mathfrak{S}_{N+l}}{\underset{\rho\in\mathfrak{S}_{N+k}}{\sum}}\sign(\rho)\sign(\sigma)\det[r_{a\rho(b)}]\underset{1\leq a,b\leq k}{\ }\det[s_{a\sigma(b)}]\underset{1\leq a,b\leq l}{\ }\times\nonumber\\
 \fl &\times&\int\limits_{\mathbb{C}^N}\det[R_{\rho(b)}(z_a,z_a^*)]\underset{k+1\leq b\leq N+k}{\underset{1\leq a\leq N}{\ }}\det[S_{\sigma(b)}(z_a,z_a^*)]\underset{l+1\leq b\leq N+l}{\underset{1\leq a\leq N}{\ }}d[z]\,.\label{a3.2}
\end{eqnarray}
We apply Andr\'{e}ief's integration theorem for determinants \cite{And1883} and obtain
\begin{eqnarray}
 \fl \mathcal{I}&=&\frac{N!}{k!(N-k)!l!(N-l)!}\underset{\sigma\in\mathfrak{S}_{N+l}}{\underset{\rho\in\mathfrak{S}_{N+k}}{\sum}}\sign(\rho)\sign(\sigma)\det[r_{a\rho(b)}]\underset{1\leq a,b\leq k}{\ }\det[s_{a\sigma(b)}]\underset{1\leq a,b\leq l}{\ }\times\nonumber\\
 \fl &\times&\det\left[\int\limits_{\mathbb{C}}R_{\rho(a)}(z,z^*)S_{\sigma(b)}(z,z^*)d^2z\right]\underset{l+1\leq b\leq N+l}{\underset{k+1\leq a\leq N+k}{\ }}\,.\label{a3.3}
\end{eqnarray}
This expression is an expansion of a determinant of a $(N+k+l)\times(N+k+l)$ matrix in the first $k$ columns and the first $l$ rows. We find the final result
\begin{eqnarray}
 \fl \mathcal{I}&=&(-1)^{kl}N!\det\left[\begin{array}{cc} 0 & \{s_{ab}\}\underset{1\leq b\leq N+l}{\underset{1\leq a\leq l}{\ }} \\  \{r_{ba}\}\underset{1\leq b\leq k}{\underset{1\leq a\leq N+k}{\ }} & \left\{\int\limits_{\mathbb{C}}R_{a}(z,z^*)S_{b}(z,z^*)d^2z\right\}\underset{1\leq b\leq N+l}{\underset{1\leq a\leq N+k}{\ }} \end{array}\right]\,.\label{a3.4}
\end{eqnarray}
For $k=l=0$, we, indeed, obtain the original integral theorem by Andr\'{e}ief.

\subsection{Extension of de Bruijn's integral theorem}\label{app3.2}

Consider the integral
\begin{equation}\label{a3.5}
 \fl \mathcal{J}=\int\limits_{\mathbb{C}^N}\det\left[\begin{array}{ccc} \{A_{ab}\}\underset{1\leq b\leq l}{\underset{1\leq a\leq 2N+l}{\ }} & \{B_a(z_b,z_b^*)\}\underset{1\leq b\leq N}{\underset{1\leq a\leq 2N+l}{\ }} & \{C_a(z_b,z_b^*)\}\underset{1\leq b\leq N}{\underset{1\leq a\leq 2N+l}{\ }} \end{array}\right]d[z]\,.
\end{equation}
As in \ref{app3.1}, we expand the determinant in the first $l$ columns and obtain
\begin{eqnarray}
 \fl \mathcal{J}&=&\frac{1}{(2N)!}\sum\limits_{\sigma\in\mathfrak{S}_{2N+l}}\sign(\sigma)\prod\limits_{j=1}^lA_{\sigma(j)j}\times\nonumber\\
 \fl &\times&\int\limits_{\mathbb{C}^N}\det\left[\begin{array}{cc} \{B_{\sigma(a)}(z_b,z_b^*)\}\underset{1\leq b\leq N}{\underset{l+1\leq a\leq 2N+l}{\ }} & \{C_{\sigma(a)}(z_b,z_b^*)\}\underset{1\leq b\leq N}{\underset{l+1\leq a\leq 2N+l}{\ }} \end{array}\right]d[z]\,.\label{a3.6}
\end{eqnarray}
We define the quantity
\begin{eqnarray}\label{a3.7}
 D_{ab}=\int\limits_{\mathbb{C}}\left[B_{a}(z,z^*)C_{b}(z,z^*)-B_{b}(z,z^*)C_{a}(z,z^*)\right]d[z]\,.
\end{eqnarray}
Then, we apply the original version of de Bruijn's integral theorem \cite{Bru55} and find
\begin{eqnarray}
 \fl \mathcal{J}&=&\frac{(-1)^{N(N-1)/2}N!}{(2N)!}\sum\limits_{\sigma\in\mathfrak{S}_{2N+l}}\sign(\sigma)\prod\limits_{j=1}^lA_{\sigma(j)j}\Pf\left[D_{\sigma(a)\sigma(b)}\right]_{l+1\leq a,b\leq 2N+l}\,.\label{a3.8}
\end{eqnarray}
Summarizing all terms, the integral $\mathcal{J}$ is up to a constant
\begin{eqnarray}
 \fl \mathcal{J}&\sim&\Pf\left[\begin{array}{cc} 0 & \{A_{ba}\}\underset{1\leq b\leq 2N+l}{\underset{1\leq a\leq l}{\ }} \\ \{-A_{ab}\}\underset{1\leq b\leq l}{\underset{1\leq a\leq 2N+l}{\ }} & \{D_{ab}\}_{1\leq a,b\leq 2N+l} \end{array}\right]\,.\label{a3.9}
\end{eqnarray}
We fix the constant by the particular choice
\begin{equation}\label{a3.10}
 [-A_{ab}]\underset{1\leq b\leq l}{\underset{1\leq a\leq 2N+l}{\ }}=\left[\begin{array}{c} \eins_l \\ 0 \end{array}\right]
\end{equation}
which yields
\begin{eqnarray}
 \fl \mathcal{J}&=&(-1)^{N(N-1)/2+l(l-1)/2}N!\Pf\left[\begin{array}{cc} 0 & \{A_{ba}\}\underset{1\leq b\leq 2N+l}{\underset{1\leq a\leq l}{\ }} \\ \{-A_{ab}\}\underset{1\leq b\leq l}{\underset{1\leq a\leq 2N+l}{\ }} & \{D_{ab}\}_{1\leq a,b\leq 2N+l} \end{array}\right]\,.\label{a3.11}
\end{eqnarray}
For $l=0$, this is indeed de Bruijn's integral theorem.

\section{Calculating integrals of squared--Vandermonde type}\label{app4}

We derive the cases $(k_1-k_2)=(l_1-l_2)\leq N$ and $(k_2-k_1),(l_2-l_1)\leq N$ in \ref{app4.1} and \ref{app4.2}, respectively.

\subsection{The case $(k_1-k_2)=(l_1-l_2)\leq N$}\label{app4.1}

With help of Eq.~\eref{2.12}, we rewrite the integrand \eref{3.17} as a product of two determinants
\begin{eqnarray}
 \fl&&\int\limits_{\mathbb{C}^{N}}\prod\limits_{j=1}^{N}g(z_{j}) \sqrt{\Ber_{k_1/k_2+N}^{(2)}(\tilde{z})}\sqrt{\Ber_{l_1/l_2+N}^{(2)}(\hat{z})}d[z]=\nonumber\\
 \fl&=&(-1)^{(l_1-k_1)(l_1+k_1-1)/2}\int\limits_{\mathbb{C}^{N}}\prod\limits_{j=1}^{N}g(z_{j}) \det\left[\begin{array}{cc}\left\{\displaystyle\frac{1}{\kappa_{a1}-\kappa_{b2}}\right\}\underset{1\leq b\leq k_2}{\underset{1\leq a\leq k_1}{\ }} & \left\{\displaystyle\frac{1}{\kappa_{a1}-z_b}\right\}\underset{1\leq b\leq N}{\underset{1\leq a\leq k_1}{\ }} \\ \left\{\kappa_{b2}^{a-1}\right\}\underset{1\leq b\leq k_2}{\underset{1\leq a\leq d}{\ }} & \left\{z_b^{a-1}\right\}\underset{1\leq b\leq N}{\underset{1\leq a\leq d}{\ }} \end{array}\right]\times\nonumber\\
 \fl&\times&\det\left[\begin{array}{cc}\left\{\displaystyle\frac{1}{\lambda_{a1}-\lambda_{b2}}\right\}\underset{1\leq b\leq l_2}{\underset{1\leq a\leq l_1}{\ }} & \left\{\displaystyle\frac{1}{\lambda_{a1}-z_{b}^*}\right\}\underset{1\leq b\leq N}{\underset{1\leq a\leq l_1}{\ }} \\ \left\{\lambda_{b2}^{a-1}\right\}\underset{1\leq b\leq l_2}{\underset{1\leq a\leq d}{\ }} & \left\{z_{b}^{*\,a-1}\right\}\underset{1\leq b\leq N}{\underset{1\leq a\leq d}{\ }} \end{array}\right]d[z]\,.\label{a4.1}
\end{eqnarray}
Using the definitions \eref{3.20}-\eref{3.25}, we apply the theorem of \ref{app3.1} and find
\begin{eqnarray}
 \fl&&\int\limits_{\mathbb{C}^{N}}\prod\limits_{j=1}^{N}g(z_{j}) \sqrt{\Ber_{k_1/k_2+N}^{(2)}(\tilde{z})}\sqrt{\Ber_{l_1/l_2+N}^{(2)}(\hat{z})}d[z]=(-1)^{(l_2+k_2)(l_1+k_1-1)/2}N!\times\nonumber\\
 \fl&\times&\det\left[\begin{array}{ccc} 0 & \left\{\displaystyle\frac{1}{\lambda_{b1}-\lambda_{a2}}\right\}\underset{1\leq b\leq l_1}{\underset{1\leq a\leq l_2}{\ }} & \left\{\mathbf{\Lambda}_{d}(\lambda_{a2})\right\}\underset{1\leq a\leq l_2}{\ } \\  \left\{\displaystyle\frac{1}{\kappa_{a1}-\kappa_{b2}}\right\}\underset{1\leq b\leq k_2}{\underset{1\leq a\leq k_1}{\ }} & \left\{\widetilde{Z}_{\frac{1/0}{\overset{\ }{1/0}}}^{(1)}(\kappa_{a1},\lambda_{b1})\right\}\underset{1\leq b\leq l_1}{\underset{1\leq a\leq k_1}{\ }} & \left\{\mathbf{\widetilde{F}}_{d}(\kappa_{a1})\right\}\underset{1\leq a\leq k_1}{\ } \\ \left\{\mathbf{K}_{d}(\kappa_{b2})\right\}\underset{1\leq b\leq k_2}{\ } & \left\{\mathbf{\widetilde{F}^{\,(*)}}_{d}(\lambda_{b1})\right\}\underset{1\leq b\leq l_1}{\ } & \mathbf{\widetilde{M}}_{d} \end{array}\right]\,.\label{a4.2}
\end{eqnarray}
The last step is the same as in \ref{app2.1}. We separate the matrix $\mathbf{\widetilde{M}}_{d}$ from the determinant by inverting it. This yields Eq.~\eref{3.19}.

\subsection{The case $(k_2-k_1),(l_2-l_1)\leq N$}\label{app4.2}

We consider the integral
\begin{eqnarray}
 \fl&&\int\limits_{\mathbb{C}^{N}}\prod\limits_{j=1}^{N}g(z_{j}) \sqrt{\Ber_{k_1/k_2+N}^{(2)}(\tilde{z})}\sqrt{\Ber_{l_1/l_2+N}^{(2)}(\hat{z})}d[z]=(-1)^{l_1(l_1-1)/2+k_1(k_1-1)/2}\times\nonumber\\
 \fl&\times&\int\limits_{\mathbb{C}^{N}}\prod\limits_{j=1}^{N}g(z_{j}) \det\left[\begin{array}{c}\left\{\displaystyle\frac{1}{\kappa_{b1}-\kappa_{a2}}\right\}\underset{1\leq b\leq k_1}{\underset{1\leq a\leq k_2}{\ }} \\ \left\{\displaystyle\frac{1}{\kappa_{b1}-z_a}\right\}\underset{1\leq b\leq k_1}{\underset{1\leq a\leq N}{\ }} \\ \left\{\kappa_{b1}^{a-1}\right\}\underset{1\leq b\leq k_1}{\underset{1\leq a\leq d_{\kappa}}{\ }} \end{array}\right]\det\left[\begin{array}{c}\left\{\displaystyle\frac{1}{\lambda_{b1}-\lambda_{a2}}\right\}\underset{1\leq b\leq l_1}{\underset{1\leq a\leq l_2}{\ }} \\ \left\{\displaystyle\frac{1}{\lambda_{b1}-z_a^*}\right\}\underset{1\leq b\leq l_1}{\underset{1\leq a\leq N}{\ }} \\ \left\{\lambda_{b1}^{a-1}\right\}\underset{1\leq b\leq l_1}{\underset{1\leq a\leq d_{\lambda}}{\ }} \end{array}\right]d[z]\,.\label{a4.3}
\end{eqnarray}
Using the result of \ref{app3.1} we get
\begin{eqnarray}
 \fl&&\int\limits_{\mathbb{C}^{N}}\prod\limits_{j=1}^{N}g(z_{j}) \sqrt{\Ber_{k_1/k_2+N}^{(2)}(\tilde{z})}\sqrt{\Ber_{l_1/l_2+N}^{(2)}(\hat{z})}d[z]=(-1)^{(l_1+k_1)(l_1+k_1-1)/2+N(k_2+l_2+1)}\times\nonumber\\
 \fl&\times&N!\det\left[\begin{array}{ccc} 0 & 0 & \left\{\displaystyle\frac{1}{\lambda_{b1}-\lambda_{a2}}\right\}\underset{1\leq b\leq l_1}{\underset{1\leq a\leq l_2}{\ }} \\ 0 & 0 & \left\{\lambda_{b1}^{a-1}\right\}\underset{1\leq b\leq l_1}{\underset{1\leq a\leq d_{\lambda}}{\ }} \\ \left\{\displaystyle\frac{1}{\kappa_{a1}-\kappa_{b2}}\right\}\underset{1\leq b\leq k_2}{\underset{1\leq a\leq k_1}{\ }} &  \left\{\kappa_{a1}^{b-1}\right\}\underset{1\leq b\leq d_{\kappa}}{\underset{1\leq a\leq k_1}{\ }} & \left\{\widetilde{Z}_{\frac{1/0}{\overset{}{1/0}}}^{(1)}(\kappa_{a1},\lambda_{b1})\right\}\underset{1\leq b\leq l_1}{\underset{1\leq a\leq k_1}{\ }} \end{array}\right]\,,\label{a4.4}
\end{eqnarray}
which is the desired formula.

\section{Calculation of the flat Fourier transform in Eq.~\eref{5.14}}\label{app5}

We consider the flat Fourier transform
\begin{eqnarray}
 \fl J &=& \int\limits_{\mathbb{R}^{2k}}\prod\limits_{a=1}^k\prod\limits_{b=1}^N\frac{e^{-\imath\psi}s_{a2}+\imath\varepsilon+\alpha E_{b}^{(0)}}{s_{a1}+\imath\varepsilon+\alpha E_{b}^{(0)}}\exp\left[-\imath\Str rs^+\right]\sqrt{{\rm Ber}_{k/k}^{(2)}(s)}d[s]\,. \label{a5.1}
\end{eqnarray}
By extending this integral with a Vandermonde determinant of $-\alpha E^{(0)}$ and using Eq.~\eref{2.11}, we find the determinant
\begin{eqnarray}
 \fl J &=& \frac{(-1)^{k(k-1)/2}}{\Delta_N(\alpha E^{(0)})}\det\left[\begin{array}{cc} \left\{\displaystyle J_1(\tilde{r}_{ab})\right\}\underset{1\leq a,b\leq k}{\ } & \left\{\displaystyle J_2(r_{a1},\alpha E_b^{(0)})\right\}\underset{1\leq b\leq N}{\underset{1\leq a\leq k}{\ }} \\ \left\{\displaystyle J_{3,a}(r_{b2})\right\}\underset{1\leq b\leq k}{\underset{1\leq a\leq N}{\ }} & \left\{\left(-\alpha E_b^{(0)}\right)^{a-1}\right\}\underset{1\leq a,b\leq N}{\ } \end{array}\right]\,,\label{a5.2}
\end{eqnarray}
where $s_1^+=s_1+\imath\varepsilon$, $s_2^+=s_2+\imath e^{\imath\psi}\varepsilon$ and $\tilde{r}_{ab}=\diag(r_{a1},e^{\imath\psi}r_{b2})$. Hence, we have to calculate three types of integrals. The integrals in the off-diagonal blocks are the simpler ones. We have
\begin{eqnarray}
 \fl &&J_2(r_{a1},\alpha E_b^{(0)})=\displaystyle\int\limits_{\mathbb{R}}\frac{\exp\left[-\imath r_{a1}s_1^+\right]}{(s_1^++\alpha E_b^{(0)})}\left(\frac{-\alpha E_b^{(0)}}{s_1^+}\right)^Nds_1=\nonumber\\
 \fl&=&\displaystyle\frac{(-\imath)^{N+1}\left(-\alpha E_b^{(0)}\right)^N}{(N-1)!}\int\limits_{\mathbb{R}}\int\limits_{\mathbb{R}^+}\int\limits_{\mathbb{R}^+}t_2^{N-1}\exp\left[-\imath r_{a1}s_1^++\imath(s_1^++\alpha E_b^{(0)})t_1+\imath s_1^+t_2\right]d[t,s]=\nonumber\\
 \fl&=&\displaystyle\frac{(-\imath)^{N+1}2\pi\left(-\alpha E_b^{(0)}\right)^N}{(N-1)!}\int\limits_{\mathbb{R}^+}\int\limits_{\mathbb{R}^+}\delta(t_1+t_2-r_{a1})t_2^{N-1}\exp\left[\imath\alpha E_b^{(0)}t_1\right]dt_1dt_2=\nonumber\\
 \fl&=&\displaystyle\frac{(-\imath)^{N+1}2\pi\left(-\alpha E_b^{(0)}\right)^N}{(N-1)!}\Theta(r_{a1})\int\limits_{0}^{r_{a1}}t_2^{N-1}\exp\left[-\imath\alpha E_b^{(0)}(t_2-r_{a1})\right]dt_2=\nonumber\\
 \fl&=&\displaystyle-2\pi\imath\Theta(r_{a1})\exp\left[\imath\alpha E_b^{(0)}r_{a1}\right]+2\pi\imath\Theta(r_{a1})\sum\limits_{n=0}^{N-1}\frac{1}{n!}\left(\imath\alpha E_b^{(0)}r_{a1}\right)^{n}=\nonumber\\
 \fl&=&\displaystyle-2\pi\imath\Theta(r_{a1})\sum\limits_{n=N}^{\infty}\frac{1}{n!}\left(\imath\alpha E_b^{(0)}r_{a1}\right)^{n}\label{a5.3}
\end{eqnarray}
and
\begin{equation}\label{a5.4}
 \fl\displaystyle J_{3,a}(r_{b2})=\int\limits_{\mathbb{R}}\left(e^{-\imath\psi}s_2^+\right)^{a-1}\exp\left[\imath r_{b2}s_2^+\right]ds_2=2\pi\left(\frac{e^{-\imath\psi}}{\imath}\frac{\partial}{\partial r_{b2}}\right)^{a-1}\delta(r_{b2})
\end{equation}
The integrand of the integral
\begin{equation}
  J_1(\tilde{r}_{ab})=\displaystyle\int\limits_{\mathbb{R}^2}\frac{\exp\left[-\imath\Str\tilde{r}_{ab}s^+\right]}{s_1-e^{-\imath\psi}s_2}\left(\frac{s_2^+}{s_1^+}\right)^Nd[s]\label{a5.5}
\end{equation}
has to be interpreted as a distribution. It is up to an Efetov--Wegner term
\begin{equation}
  J_1(\tilde{r}_{ab})=\displaystyle2\pi e^{\imath\psi}\int\limits_{\Sigma_{-\psi}(1)}\exp\left[-\imath\Str\tilde{r}_{ab}\sigma^+\right]\Sdet^{-N}\sigma^+d[\sigma]+2\pi\imath\,.\label{a5.6}
\end{equation}
This is the supersymmetric Ingham-Siegel integral \cite{Guh06}. We employ the result of Refs.~\cite{Guh06,KGG08} and obtain
\begin{equation}
  J_1(\tilde{r}_{ab})=\displaystyle-2\pi \frac{r_{a1}^N\Theta(r_{a1})}{r_{a1}-e^{\imath\psi}r_{b2}}\left(-e^{-\imath\psi}\frac{\partial}{\partial r_{b2}}\right)^{N-1}\delta(r_{b2})\,.\label{a5.7}
\end{equation}
Thus, we get for the integral \eref{5.1}
\begin{eqnarray}
  \fl J &=& \det\hspace*{-0.11cm}\left[\begin{array}{cc} \left\{\displaystyle \frac{r_{a1}^N}{r_{a1}-e^{\imath\psi}r_{b2}}\left(-e^{-\imath\psi}\frac{\partial}{\partial r_{b2}}\right)^{N-1}\right\}\underset{1\leq a,b\leq k}{\ } & \left\{\displaystyle\sum\limits_{n=N}^{\infty}\frac{\imath}{n!}\left(\imath\alpha E_b^{(0)}r_{a1}\right)^{n}\right\}\underset{1\leq b\leq N}{\underset{1\leq a\leq k}{\ }} \\ \left\{\displaystyle2\pi\left(\frac{e^{-\imath\psi}}{\imath}\frac{\partial}{\partial r_{b2}}\right)^{a-1}\right\}\underset{1\leq b\leq k}{\underset{1\leq a\leq N}{\ }} & \left\{\left(-\alpha E_b^{(0)}\right)^{a-1}\right\}\underset{1\leq a,b\leq N}{\ } \end{array}\right]\hspace*{-0.12cm}\times\nonumber\\
 \fl&\times& \frac{(-2\pi)^k(-1)^{k(k-1)/2}\Theta(r_1)\delta(r_2)}{\Delta_N(\alpha E^{(0)})}\label{a5.9}\,,
\end{eqnarray}
where $\Theta(r_1)$ indicates that every bosonic eigenvalue $r_{a1}$ has to be positive definite. The distribution $\delta(r_2)$ is the product of all Dirac distributions $\delta(r_{b2})$.

\section{Laguerre ensembles in an external field}\label{app6}

We consider the Laguerre ensemble
\begin{equation}\label{a6.1}
 P_\nu(H)=\prod\limits_{j=1}^N\left[\left(\frac{c}{\pi}\right)^{j-1}\frac{c^{\nu+1}}{\Gamma(\nu+j)}\right]\exp\left(-c\tr H\right){\det}^{\nu}H\Theta(H)\,,
\end{equation}
where $\nu,c\in\mathbb{R}^+$ are some constants and $\Theta$ is the Heaviside distribution for matrices, i.e. it is unity for positive definite matrices and zero else. The characteristic function is
\begin{equation}\label{a6.2}
 \mathcal{F}P_\nu(H)=c^{(N+\nu)N}{\det}^{-N-\nu}(c\eins_N-\imath H)
\end{equation}
and the supersymmetric extension is, hence,
\begin{equation}\label{a6.3}
 \Phi_\nu(\rho)=c^{(N+\nu)(k_1-k_2)}\Sdet^{-N-\nu}(c\eins_{k_1+k_2}-\imath\rho)\,.
\end{equation}
We notice that $\Phi$ factorizes, i.e. it fulfills Eq.~\eref{5.19}. Thus we can apply the calculations in Sec.~\ref{sec5.3}.

The function $\widetilde{R}_1(x_a,x_b)$, see Eq.~\eref{5.21}, is up to the Efetov--Wegner term, which is the normalization in this case, the same as the generating function \eref{5.4} for $k=1$ and $\alpha=0$,
\begin{eqnarray}\label{a6.4}
 \fl&&\widetilde{R}_1(x_a,x_b)\sim\frac{1}{x_a-x_b}\times\\
 \fl&\times&\displaystyle\underset{\varepsilon\searrow0}{\lim}\int\limits_{\Herm(N)}P_\nu(H)\left[\frac{\det(H-x_b\eins_N)}{\det(H-(x_a-\imath\varepsilon)\eins_N)}-\frac{\det(H-x_b\eins_N)}{\det(H-(x_a+\imath\varepsilon)\eins_N)}\right]d[H]\,.\nonumber
\end{eqnarray}
Let $\pi_N^{(\nu)}$ the orthogonal polynomials of order $N$ with respect to the probability density $P_\nu$, i.e. $\pi_N^{(\nu)}(x)=x^N+\ldots$ are the associated Laguerre polynomials. Then, we find
\begin{eqnarray}\label{a6.5}
 \fl\widetilde{R}_1(x_a,x_b)&=&\frac{\pi(-\imath)^{N-1}2}{(N+\nu-1)!}\times\\
 &\times&\frac{\pi_N^{(\nu)}(cx_a)\pi_{N-1}^{(\nu)}(cx_b)-\pi_{N-1}^{(\nu)}(cx_a)\pi_N^{(\nu)}(cx_b)}{x_a-x_b}(cx_a)^\nu\exp(-cx_a)\Theta(x_a)\,,\nonumber
\end{eqnarray}
which is indeed the determinantal kernel for the case $\alpha=0$.

For calculating the second function $\widetilde{R}_2(\alpha E_b^{(0)},x_a)$, see Eq.~\eref{5.22}, we consider the integral
\begin{equation}\label{a6.6}
 \mathcal{I}_n(\alpha E_b^{(0)},x_a)=\displaystyle\imath\int\limits_{\mathbb{R}}\Phi_\nu(r_1)\exp\left[-\imath r_1x_a\right]\frac{\left(\imath\alpha E_b^{(0)}r_{1}\right)^{n}}{n!}dr_1\,.
\end{equation}
It has a structure similar to Eq.~\eref{a6.4},
\begin{eqnarray}\label{a6.7}
 \fl&&\mathcal{I}_n(\alpha E_b^{(0)},x_a)\sim\left(\alpha E_b^{(0)}\right)^{n}\times\\
 \fl&\times&\displaystyle\underset{\varepsilon\searrow0}{\lim}\int\limits_{\Herm(n+1)}P_{N+\nu-1-n}(H)\left[\frac{1}{\det(H-(x_a-\imath\varepsilon)\eins_{n+1})}-\frac{1}{\det(H-(x_a+\imath\varepsilon)\eins_{n+1})}\right]d[H]\,.\nonumber
\end{eqnarray}
Thus, we have
\begin{eqnarray}\label{a6.8}
 \fl\mathcal{I}_n(\alpha E_b^{(0)},x_a)&=&\displaystyle\frac{2\pi\imath c}{(N+\nu-1)!n!}\times\\
 \fl&\times&\left(\alpha E_b^{(0)}c\right)^{n}\pi_{n+1}^{(N+\nu-1-n)}(cx_a)(cx_a)^{N+\nu-1-n}\exp(-cx_a)\Theta(x_a)\,,\nonumber
\end{eqnarray}
which also follows by directly integrating Eq.~\eref{a6.6}. We combine this result with the definition \eref{5.22} and find
\begin{eqnarray}\label{a6.9}
 \fl&&\widetilde{R}_2(\alpha E_b^{(0)},x_a)=\\
 \fl&=&\displaystyle\left[\pi_{1}^{(N+\nu-1)}(c[x_a-\alpha E_b^{(0)}])(c[x_a-\alpha E_b^{(0)}])^{N+\nu-1}\exp(-c[x_a-\alpha E_b^{(0)}])\Theta([x_a-\alpha E_b^{(0)}])-\right.\nonumber\\
 \fl&-&\displaystyle\left.\sum\limits_{n=0}^N\frac{1}{n!}\left(\alpha E_b^{(0)}c\right)^{n}\pi_{n+1}^{(N+\nu-1-n)}(cx_a)(cx_a)^{N+\nu-1-n}\exp(-cx_a)\Theta(x_a)\right]\frac{2\pi\imath c}{(N+\nu-1)!}\,.\nonumber
\end{eqnarray}
We notice that the first term of $\widetilde{R}_2(\alpha E_b^{(0)},x_a)$ vanishes if $x_a$ is smaller than $\alpha E_b^{(0)}$.

Also for the function $\widetilde{R}_{a3}(x_b)$, see Eq.~\eref{5.23}, we find an expression of a form similar to Eq.~\eref{a6.4} and Eq.~\eref{a6.7},
\begin{eqnarray}\label{a6.10}
 &&\widetilde{R}_{a3}(x_b)\sim\displaystyle\int\limits_{\Herm(a-1)}P_{N+\nu+1-a}(H)\det(H-x_b\eins_{a-1})d[H]\,.
\end{eqnarray}
We easily see that this is
\begin{eqnarray}\label{a6.11}
 &&\widetilde{R}_{a3}(x_b)=\displaystyle2\pi c^{1-a}\pi_{a-1}^{(N+\nu+1-a)}(cx_b)\,.
\end{eqnarray}
This result can also be obtained by performing the integration of Eq.~\eref{5.23}.

We emphasis that our result of the Laguerre ensemble in the presence of an external source is different to those in Refs.~\cite{GuhWet97,DesFor06} since the coupling is different.

\section*{References}


\end{document}